\documentclass[%
 aip,
amsmath,amssymb,
 reprint,%
]{revtex4-1}

\usepackage{graphicx}% Include figure files
\usepackage{dcolumn}% Align table columns on decimal point
\usepackage{bm}% bold math
\usepackage[utf8]{inputenc}
\usepackage[T1]{fontenc}
\usepackage{mathptmx}
\usepackage{etoolbox}
\usepackage{multirow}
\usepackage{xcolor}
\usepackage{adjustbox}
\usepackage{colortbl}
\usepackage{siunitx}
\usepackage{booktabs}
\usepackage{makecell}

\makeatletter
\def\@email#1#2{%
 \endgroup
 \patchcmd{\titleblock@produce}
  {\frontmatter@RRAPformat}
  {\frontmatter@RRAPformat{\produce@RRAP{*#1\href{mailto:#2}{#2}}}\frontmatter@RRAPformat}
  {}{}
}%

\makeatother

\begin{document}

\preprint{AIP/123-QED}

\title{Elucidating Ion Capture and Transport Mechanisms of Preyssler Anions in Aqueous Solutions Using Biased MACE-Accelerated MD Simulations}

\author{Quadri O. Adewuyi}
\affiliation{Division of Energy, Matter and Systems, School of Science and Engineering, University of Missouri $-$ Kansas City, Kansas City 64110, MO, United States}
\author{Suchona Akter}
\affiliation{Division of Energy, Matter and Systems, School of Science and Engineering, University of Missouri $-$ Kansas City, Kansas City 64110, MO, United States}
\author{Md Omar Faruque}
\affiliation{Division of Energy, Matter and Systems, School of Science and Engineering, University of Missouri $-$ Kansas City, Kansas City 64110, MO, United States}
\author{Dil K. Limbu}
\affiliation{Division of Energy, Matter and Systems, School of Science and Engineering, University of Missouri $-$ Kansas City, Kansas City 64110, MO, United States}
\author{Zhonghua Peng$^*$}
\affiliation{Division of Energy, Matter and Systems, School of Science and Engineering, University of Missouri $-$ Kansas City, Kansas City 64110, MO, United States}
\email{pengz@umkc.edu}
\author{Praveen K. Thallapally$^*$}
\affiliation{Pacific Northwest National Laboratory, Richland, WA 99352, USA}
\email{praveen.thallapally@pnnl.gov}
\author{Mohammed R. Momeni$^*$}
\affiliation{Division of Energy, Matter and Systems, School of Science and Engineering, University of Missouri $-$ Kansas City, Kansas City 64110, MO, United States}
\email{mmomenitaheri@umkc.edu}
\date{\today}

\begin{abstract}
Equilibrium and biased MACE accelerated MD simulations in aqueous solutions are performed to investigate the ion capture and transport mechanisms of the \{P$_5$W$_{30}$\} Preyssler anion (PA) as the smallest representative member of the extended polyoxometalate (POM) family with an internal cavity. The unique interatomic interactions present in the internal cavity vs. exterior of PA are carefully investigated using equilibrium MACE MD simulations for two representative Na(H$_2$O)@PA and Na@PA complexes. Our careful analyses of radial distribution functions and coordination numbers show that the presence of confined water in Na(H$_2$O)@PA has profound modulating effects on the nature of the interactions of the encapsulated ion with the oxygens of the PA cavity. Using well-converged MACE-accelerated multiple walker well-tempered metadynamics simulations, two different associative and dissociative ion transport mechanisms were carefully investigated for Na$^+$ as one of the most abundant and representative ions present in seawater and saline solutions. By comparing systems with and without confined water, it was found that the presence of only one pre-encapsulated confined water in Na(H$_2$O)@PA dramatically changes the free energy landscape of ion transport processes. It was also found that the contraction and dilation of the two windows present in PA directly influence the Na$^+$ and H$_2$O transport. Results from this work are helpful as they show a viable path toward tuning the ion exchange and transport phenomena in aqueous solutions of POM molecular clusters and frameworks.
\end{abstract}

\maketitle

\section{\label{sec1:intro}INTRODUCTION}
Critical metals, including lithium, nickel, cobalt, and manganese, essential for advanced manufacturing and low-carbon technologies, are becoming scarce in land sources but abundant in saline solutions.\cite{diallo2015mining,can2021recovery} This has made extraction from aqueous sources such as saltwater and brine an intriguing research field, largely due to its significant impact on national economic development and a promising alternative to hard rock mining with lower environmental impacts.\cite{kiprono2023state,loganathan2017mining} Many saline brine solutions contain substantial amounts of critical metals, which could become pollutants if not properly managed, but are valuable commodities when recovered. However, extracting these metals from saline sources presents significant challenges due to their low concentrations and the need for selective separation.\cite{buonomenna2022mining,pramanik2020extraction} 
Current extraction methods include precipitation, ion exchange, solvent extraction, sorption, and membrane-based technologies.\cite{pramanik2020extraction,loganathan2017mining} While conventional desalination membranes lack selectivity, emerging sub-nanostructured membranes and sorption technologies offer efficiency, scalability, and low cost, particularly for environmental remediation and selective metal recovery.\cite{buonomenna2022mining,edebali2019advanced,sikdar2001separation}

\begin{figure}[!t]
\includegraphics [width=0.99\linewidth]{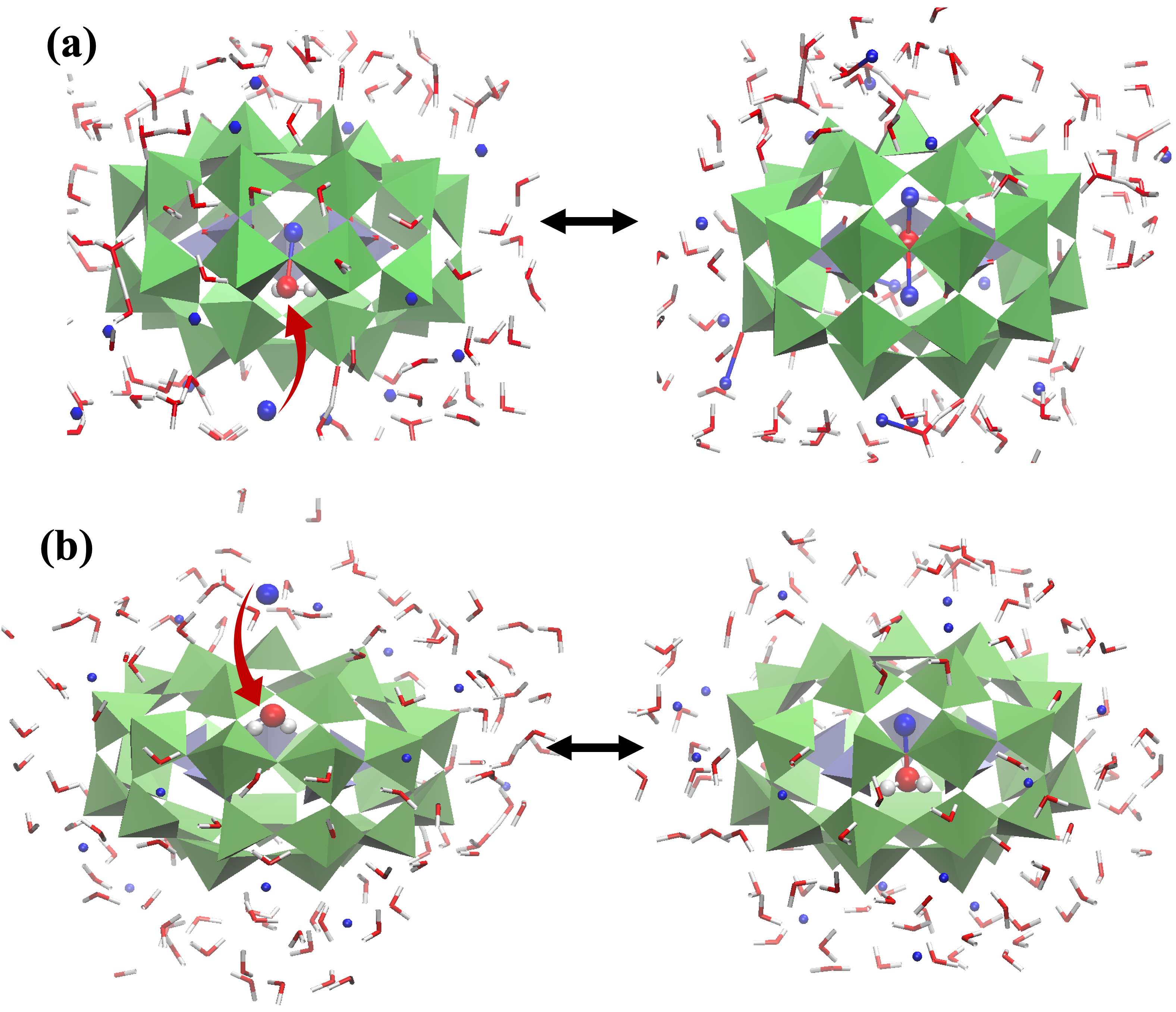}
  \caption{Ion capture mechanisms via (a) associative and (b) dissociative pathways using Preyssler anion polyoxometalates in aqueous solutions. H: white, O: red, Na$^+$: blue, PO$_4$: ice blue, WO$_6$: lime.}
    \label{fig1}
\end{figure}
Contemporary adsorbents recently employed for aqueous critical metal recovery include porous organic cages,\cite{florek2024role,iftekhar2022porous} metal-organic frameworks (MOFs),\cite{furukawa2013chemistry,yang2019metal} covalent-organic frameworks (COFs),\cite{cote2005porous,mk2024covalent,guo2024covalent} and porous polyoxometalate frameworks (POMFs),\cite{miras2012engineering,iftekhar2022porous} to name a few.
Despite advances in sorbent materials, challenges remain, including low selectivity, capacity, adsorption/desorption rates, and low recovery. These issues arise from the difficulty of selectively adsorbing trace metal ions in complex solutions with higher concentrations of competing ions present, as well as a limited understanding of the molecular-level mechanisms driving the adsorption/desorption.

This work aims to provide a mechanistic molecular-level understanding of the adsorption and release of Na$^+$ ions as the most abundant species commonly found in aqueous saline and brine solutions using robust, all-inorganic polyoxometalate (POM) molecular clusters. Redox-active POMs are nanosized anionic metal-oxides of early transition metals. They can effectively transport protons by dispersing negative charges across the surface oxygen atoms, lowering their effective surface charge density.\cite{niinomi2017high,uchida2019frontiers,iwano2024tuning} 
Their high thermal stability, flexible binding modes, and stability over a wide pH range of 1-10 make POMs ideal candidates for selective adsorptive separation in harsh aqueous environments.\cite{dey2024preyssler,boskovic2017rare,han2022cluster} 
%MACE are graph-based equivariant message passing neural networks (MPNNs) based on atomic cluster expansion(ACE).\cite{batatia2022mace,batatia2023foundation}
Specifically, the phosphotungstate Preyssler anion (PA, [X$^{n+}$P$_5$W$_{30}$O$_{110}$]$^{(15-n)-}$ with X = metal ions) with a porous doughnut-shaped structure as the smallest member of the extended POM family with an internal cavity of $\approx$5~\AA\cite{dey2024preyssler,lu2007new,alizadeh1985heteropolyanion,kim1999slow,fernandez2007polyoxometalates} is an ideal sorbent for selectively adsorbing (mono-)hydrated metal ions of radius $\approx$4~\AA~(Fig. \ref{fig1}).

PA has been experimentally shown to encapsulate potassium \cite{hayashi2018preparation,hayashi2016encapsulation} and sodium\cite{wihadi2020synthesis} ions in its central cavity. Composed of five PW$_6$  units, each consisting of two groups of three corner-sharing WO$_6$ octahedra, PA has an ellipsoid shape and an internal fivefold symmetry axis, with its doughnut shape believed to facilitate ion adsorption/desorption from aqueous solutions. 
Interestingly, the encapsulated metal ion is shown to significantly affect the POM's charge density, redox behavior, and basicity.\cite{fernandez2007polyoxometalates,creaser1993rigid,antonio1994cerium,soderholm1995coordination,antonio1996redox,antonio1997implications,antonio1999redox,williams2000formation,fernandez2007polyoxometalates,cardona2012lanthanoid,takahashi2014preparation} Examples of experimentally reported encapsulated ions include Na$^+$,\cite{alizadeh1985heteropolyanion, creaser1993rigid, kim1999slow} K$^+$,\cite{zhao2014anion,hu2015four, hayashi2018preparation} Ag$^+$,\cite{liang2014solvothermal,du2017two, kato2017synthesis} lanthanids,\cite{creaser1993rigid,antonio1994cerium,dickman1996structures,williams2000formation,granadeiro2013lanthanopolyoxometalates,takahashi2014preparation} Ca$^{2+}$,\cite{creaser1993rigid,kim1999slow,takahashi2014preparation} Bi$^{3+}$,\cite{creaser1993rigid,takahashi2014preparation,hayashi2015cation} Y$^{3+}$,\cite{creaser1993rigid,kim1999slow,takahashi2014preparation} and actinides\cite{creaser1993rigid,dickman1996structures,kim1999slow,williams2000formation,antonio2008stabilization}. 

While the extended family of POMs has been extensively studied for catalytic applications,\cite{zhang2023state,zhang2021polyoxometalate,miras2014polyoxometalate,streb2012new,dolbecq2010hybrid,alizadeh2003novel} their structure-property relationships as sorbents for different critical metal capture and separation in aqueous solutions remain underexplored. This work addresses this knowledge gap by providing an in-depth exploration of ion capture and transport mechanisms in aqueous solutions of PA using equilibrium and well-tempered metadynamics (WT-MetaD) simulations enabled by accurate multi atomic cluster expansion (MACE)\cite{batatia2022mace,batatia2023foundation} foundation models. The accuracy of atomistic simulations relies on capturing interatomic interactions with high accuracy. While classical force fields are efficient for large-scale simulations, they often lack accuracy, have limited transferability, and are nonreactive by construction (with the exception of ReaxFF).\cite{hunenberger1999empirical,amira2005car,beret2008explaining,senftle2016reaxff,dauber2019biomolecular} \textit{Ab initio} MD (AIMD) simulations improve accuracy by calculating forces on-the-fly via accurate electronic structure methods but suffer from limited length/time scales due to their high computational cost. Recently, machine learning potentials (MLPs) have emerged as a promising alternative, providing a balance between accuracy and efficiency.\cite{bonati2018silicon, zhao2019commensurate, niu2020ab,batatia2022mace,kovacs2023evaluation,kovacs2023mace,chernyshov2024mace, jctc_20_164,brookes2025co2, behler2016perspective,butler2018machine,deringer2019machine,kang2020large,mishin2021machine,behler2021four,schran2021machine}
MACE-MLPs, trained on extended datasets of material properties, bypass the cost of \textit{ab initio} methods, enabling large-scale simulations with excellent performance for diverse applications.\cite{kovacs2023mace,jctc_20_164,chernyshov2024mace,brookes2025co2}
This study utilizes MACE foundation models for both equilibrium and biased WT-MetaD simulations to investigate the capture and transport mechanisms of the targeted hydrated ions and PA. Two different ``associative'' and ``dissociative'' ion capture mechanisms are carefully investigated, with the effects of the presence of confined water in the PA cavity analyzed. This work is structured as follows. Section II provides theory and simulation details. The results and discussion are provided in Section III. Finally, conclusions and future outlooks are provided in the Conclusions section.

\section{\label{sec2:methods} Simulation Details}
\begin{figure}[!t]
\includegraphics [width=0.9\linewidth]{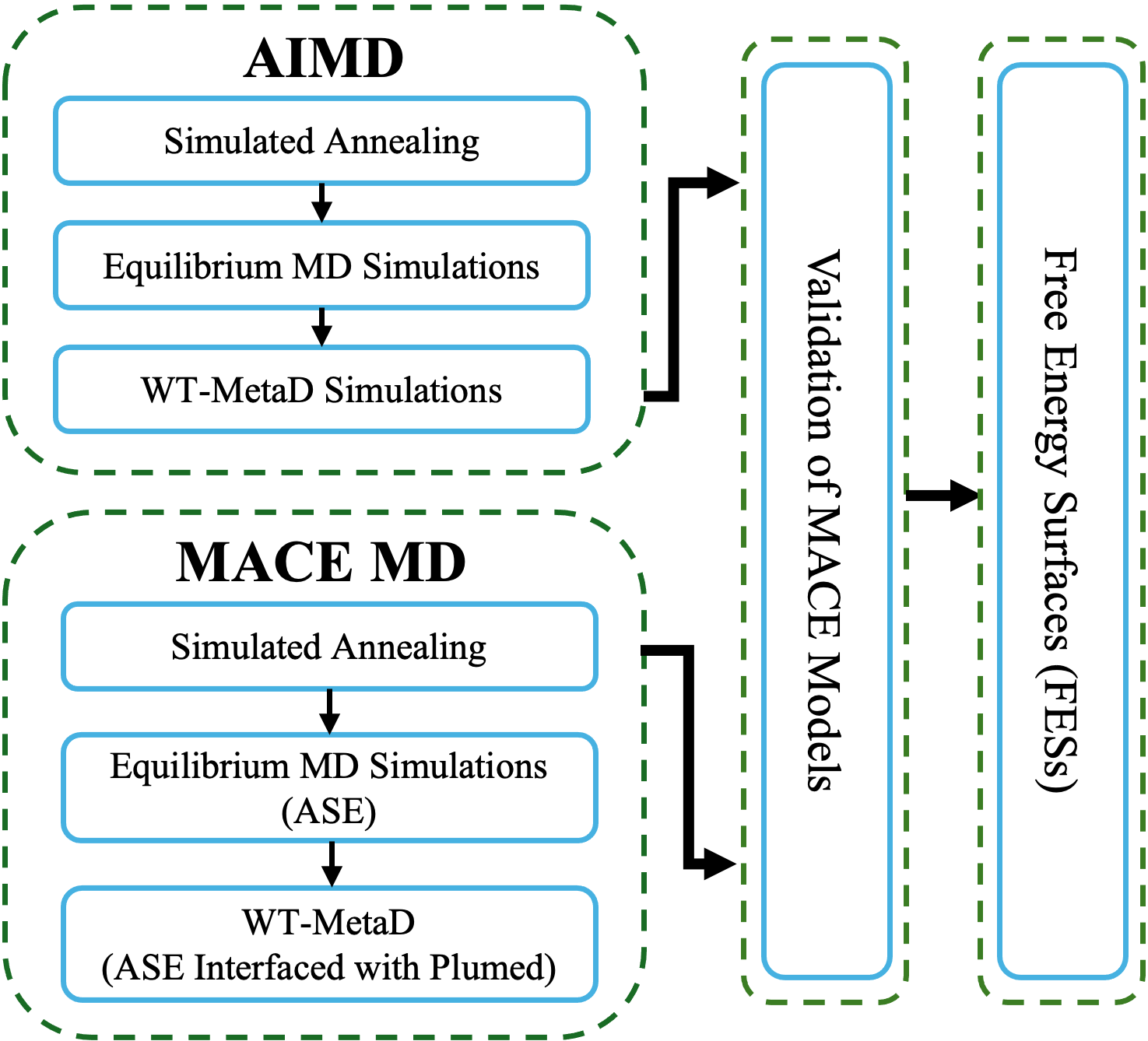}
    \caption{The simulation workflow adapted in this work.}
    \label{fig2}
\end{figure}

The adapted workflow combining electronic structure calculations, AIMD, and MACE MD simulations for studying ion transport in aqueous solutions of PA is shown in Fig. \ref{fig2}. The isolated PA is fully minimized and solvated with water and 15 ions using PACKMOL\cite{martinez2009packmol}. Both AIMD and MACE MD simulations involving simulated annealing, equilibrium MD simulations, and WT-MetaD simulations are then performed to explore the free-energy landscape of ion capture by incorporating thermal energy and conformational entropy from MD simulations at room temperature. AIMD is used as a reference, while MACE MD, interfaced with ASE and PLUMED, is used for generating well-converged simulated data. The free energy surfaces (FES) from MACE MD are benchmarked against AIMD for validation. More details are provided below.

\subsection{\label{sec21:AIMD}Electronic Structure Calculations}
All electronic structure calculations were performed using the revised PBE (revPBE)\cite{perdew1996generalized} functional within the periodic boundary conditions as implemented in the QUICKSTEP module of CP2K.\cite{vandevondele2005quickstep,hutter2014cp2k} To account for van der Waals interactions, Grimme's D3 dispersion correction (revPBE-D3) with Becke-Johnson
(BJ) damping was employed.\cite{grimme2010consistent,grimme2011effect} The Kohn-Sham orbitals were expanded using the double-$\zeta$ valence with polarization Gaussian basis sets (DZVP-MOLOPT-SR-GTH), while core electrons were treated with the Goedecker-Teter-Hutter pseudopotentials optimized for PBE (GTH-PBE).\cite{goedecker1996separable} The SCF convergence was set to 10$^{-5}$. The simulation cell was prepared as follows: a single isolated PA cluster was extracted from the experimental crystal structure as obtained from Ref. \citenum{kim1999slow} by removing solvent (water) molecules and ions. The cluster was then placed at the center of a cubic box with a Na--OH$_2$ molecule added to the center of its cavity, resulting in an overall stoichiometry of (Na(H$_2$O)P$_5$W$_{30}$O$_{110}$)$^{14-}$ for the NaH$_2$O@PA system (see Fig. \ref{fig1}). The choice of addition of M-H$_2$O (M = Na) to the cavity is motivated by the experimentally reported crystal structure of PA, which contains an encapsulated Na$^+$ ion bonded to a water molecule inside its cavity.\cite{kim1999slow}
First, a thorough benchmark of atomic positions and cell vectors was carried out without imposing any constraints. The structure was fully minimized at the revPBE-D3 level with plane-wave cutoffs ranging from 500-700 Ry, with/without effective Hubbard (U$_{eff}$)\cite{moore2024high} for tungsten (W) and oxygen (O) atoms. The results of the benchmarks are provided in Supplementary Material Table S1. The obtained structural parameters were found to be marginally affected by variations in the details of the minimizations and were found to agree well with the experiment. The fully minimized NaH$_2$O@PA structure with an energy cut-off of 700Ry without (U$_{eff}$) was used for MD simulations as explained in the next section.
\subsection{\label{sec21:MD}AIMD Simulations} 
 To arrive at a neutral unit cell for PA containing NaH$_2$O@PA, 14 Na$^+$ ions were randomly added using PACKMOL\cite{martinez2009packmol}. To model the aqueous solution, two different water contents were considered in this work, one with 109 and another with 149 water molecules randomly added to the exterior of PA, totaling 110 and 150H$_2$O molecules, respectively (Fig. \ref{fig1}). The lengths of the box were adjusted to achieve a density of 0.997 g/cm$^3$ for the liquid bulk water at room temperature, with lengths evaluated as $l_{x,y,z}$ = 25.07~\AA~and 25.65~\AA, for 110 and 150 water systems, respectively. For both water contents, three different plane-wave cut-offs of 300, 400, and 500 Ry were considered. To ensure proper randomizations, simulated annealing was performed in the canonical (NVT) ensemble at 700K, 500K, and 298K, 2 ps each, totaling 6 ps for all considered systems. 
 
A timestep of 1 fs was used using the CSVR thermostat\cite{bussi2007canonical} with a 20 fs time constant. Final snapshots from the 298K simulated annealing stage were used to perform production simulations in the NVT ensemble for 10 ps at 298K for each system. The first 8 ps was discarded as equilibration, with the last 2 ps used for statistical analyses. Periodic boundary conditions were employed throughout, except during the initial optimization of the single isolated PA. Considering the agreements between calculated radial distribution functions (RDFs) and the cost of on-the-fly AIMD simulations (see Supplementary Material Figs. S1 and S2), the smaller plane wave cut-off of 300Ry with a REL-CUTOFF of 50Ry was employed for 5 walker AIMD WT-MetaD simulations using the smaller 110 water system; more details are provided in the next section. \\

\subsection{\label{sec21:Benchmarking} Benchmarking Equilibrium MACE MD Simulations}
The rather large size of the considered solvated PA systems (>500 atoms) limits the application of on-the-fly AIMD simulations in calculating well-converged properties. MACE foundation models trained on large datasets of material properties provide an attractive alternative. Here, using the Atomic Simulation Environment (ASE) Python package,\cite{larsen2017atomic} MACE MD simulations were performed using the MACE-MPA-0\_{medium} model. This model was chosen for its accuracy, enabled by training on an expanded dataset (see below for details of our benchmarks).\cite{batatia2023foundation} First, using the 110H$_2$O system, simulated annealing was performed at 700K, 500K, and 298K temperatures in the NVT ensemble for a total of 150 ps, with 50 ps at each temperature. A timestep of 1 fs was used, and a Langevin thermostat with a friction coefficient of 0.1 ps$^{-1}$ was applied. Final snapshots from the simulated annealing at 298K were used as initial configurations for production simulations in the NVT ensemble, for a total of 1 ns at 298K. In the production simulations, the first 800 ps were discarded as equilibration, and the last 200 ps were used for statistical analyses.

To assess the sensitivity of the simulated data to different MACE models, extensive benchmarks using MD simulations in the NVT ensemble were performed. All simulations were run for a duration of 200 ps at 298K. The primary goal of these simulations was to evaluate the performance and accuracy of all available MACE models for capturing interatomic interactions of the PA system in aqueous solutions. Six MACE models known for their efficiency and accuracy were chosen, including: MACE-MPA-0\_{medium}, MACE-MP-0b\_{medium}, MACE-MP-0b2\_{small}, MACE-MP-0b2\_{medium}, MACE-MP-0b2\_{large}, and MACE-MP-0b3\_{medium}. Calculated RDFs were used to assess how well different MACE models are able to capture key features of the ion-water vs. ion-PA interactions compared to the reference AIMD data (see Supplementary Material Fig. S5). 
The intramolecular W-O$_{PA}$ (i.e., tungsten to oxygen in PA) RDFs from different MACE models are well aligned with one another and with AIMD. The three peaks represent W interactions with three distinct PA oxygen types, namely: (i) terminal oxygen, which shows the shortest W-O$_{PA}$ distance of 1.75~\AA\, in MACE vs. 1.70~\AA~in AIMD; (ii) bridging oxygen peaks near $\approx$ 1.95~\AA~in both MACE and AIMD; and finally (iii) PO$_4$ templating anion oxygen peaks at around 2.3~\AA~in MACE, while AIMD shows a less pronounced peak around 2.4~\AA. 

All studied MACE models are able to capture the first solvation shell of Na-O$_{PA}$ but slightly overestimate the second peak. For O$_W$-O$_W$ (oxygen of water) RDFs, both AIMD and MACE agree on peak positions, though MACE predicts slightly more structured water. 
%The discrepancy may stem from how MACE and AIMD differ in capturing electronic interactions, polarization effects, or solvent-mediated interactions. 
MACE models accurately reproduce RDFs for Na-O$_W$, O$_W$-H$_W$, H$_W$-H$_W$, and H$_W$-O$_{PA}$ pairs (Supplementary Material Figs. S5 and S6). Based on these extensive benchmarks, the MACE-MPA-0\_{medium} model was selected and used throughout this work.

\subsection{\label{sec22:wtmetad}Multiple Walker MACE Well-Tempered Metadynamics Simulations} 

Metadynamics is a well-established enhanced sampling technique that uses a set of pre-defined collective variables (CVs) represented by s = s(R) as a function of nuclear coordinates R, to build a history-dependent bias potential V(s) against previously visited configurations.\cite{laio2002escaping,valsson2016enhancing,bussi2020exploring} This approach is commonly used for accelerating the sampling of rare events, allowing large energy barriers to be surmounted within a feasible simulation time scale. The FES can then be estimated from the added history-dependent bias potential in the form of Gaussian functions,\cite{yang2022using}\\
\begin{equation}
	G(s,s_k) = \sum_{k=1} ^n W e ^{- \frac {||s-s_{k}|| ^ 2} {2 \sigma^2}},
\label{eq1}
\end{equation}
with $W$ and $\sigma$ corresponding to the height and width of the Gaussian, respectively. In this work, the well-tempered variation of metadynamics (WT-MetaD) coupled with the multiple-walkers approach\cite{raiteri2006efficient} is employed for all enhanced sampling simulations. In this approach, the height of the deposited Gaussian in Eq. \ref{eq1} is decreased exponentially as the external bias potential is deposited periodically,\cite{barducci2008well,dama2014well}
\begin{equation}
	V_n(s) = \sum_{k=1} ^n G(s,s_k) e^{[-\frac{\beta}{\gamma-1}V_{k-1}(s_{k})]}
\end{equation}
where $\beta = \frac{1}{k_BT}$ is the reciprocal temperature, and  $\gamma > 1$  is the bias factor.\\
This method requires a careful selection of CVs that can effectively capture the desired reaction mechanism. In this study, we considered two associative and dissociative mechanisms (see Fig. \ref{fig1}). In the former associative mechanism, the captured ion is exchanged with another ion from the solution, while in the latter dissociative mechanism, the ion is ejected by simple heating without another ion entering. To explore the reaction pathway for the associative mechanism, two coordination number (CN) based CVs were considered, as described in eqn.~\ref{eq:cv1a}, 

\begin{equation}
    S = \sum_{i,j} \left\{\frac{1-(\frac{r_{ij}}{r_c})^{8}}{1-(\frac{r_{ij}}{r_c})^{16}}\right\}
    \label{eq:cv1a}
\end{equation}

\noindent where r$_{ij}$ is the distance between atoms $i$ and $j$ and $r_c$ is the cutoff distance.
For the associative mechanism, the first CV tracks the coordination between the encapsulated Na$^+$ ion and free H$_2$O molecules at the exterior of the PA, with a $r_c$ set to 3.0~\AA, while the second CV corresponds to the coordination between free Na$^+$ ions at the exterior of the cavity and the oxygen atoms of the PO$_4$ units, with an $r_c$ of 3.2~\AA. The cutoff distance r$_c$ and the switching functions were fitted based on the first solvation shell peaks from the RDF analyses of the final 200 ps of the equilibrium MACE MD simulations (see Supplementary Material Fig. S10). As mentioned, these CVs allow tracking the simultaneous transport of an encapsulated Na$^+$ ion out of the PA cavity and the adsorption of a free Na$^+$ ion from the solution, enabling estimation of the overall free energy barrier for this ion exchange process.\\ 

Similarly, to explore the reaction pathway for the considered dissociative mechanism, two CN-based CVs were selected (eq.~\ref{eq:cv1a}).
The first CV tracks the coordination between the encapsulated Na$^+$ ion and the oxygens of the PO$_4$ units with the r$_c$ set to 3.2~\AA, while the second CV corresponds to the coordination between the encapsulated Na$^+$ ion and free H$_2$O molecules at the exterior of the PA, with the r$_c$ set to 3.0~\AA~(See the Supplementary Material Fig. S9). 
The two selected CVs enable analysis of the encapsulated Na$^+$ ion transport from the PA cavity to the aqueous solution by estimating its corresponding free energy barrier.
All multiple walker WT-MetaD simulations were performed using the benchmarked MACE-MPA-0\_{medium} model. Gaussian hills (1 kJ/mol height, 0.5~\AA~width) were added every 500 MD steps. All WT-MetaD simulations for MACE were performed using ASE\cite{larsen2017atomic} interfaced with PLUMED\cite{tribello2014plumed,tribello2024plumed}. Five walkers were used for all MACE simulations, each run for 250 ps, totaling 1.25 ns of simulation time.\\ 

For further validation, representative WT-MetaD runs were also performed at the reference rev-PBE-D3 level for the Na-H$_2$O@PA system using CP2K interfaced with PLUMED’s multiple-walker algorithm.\cite{tribello2014plumed,tribello2024plumed} Gaussian width and height parameters were kept the same as in the MACE simulations. Due to AIMD's high cost, each walker was run for 15 ps, totaling 75 ps across five walkers, with Gaussians added every 40 MD steps. These benchmarks were performed for ejecting the Na$^+$ ion through the considered dissociative mechanism for the Na(H$_2$O)@PA system, which resulted in a similar free energy barrier (Supplementary Material Figs. S3 and S4). More details are provided in the next sections.

\begin{figure*}
\includegraphics [width=\textwidth]{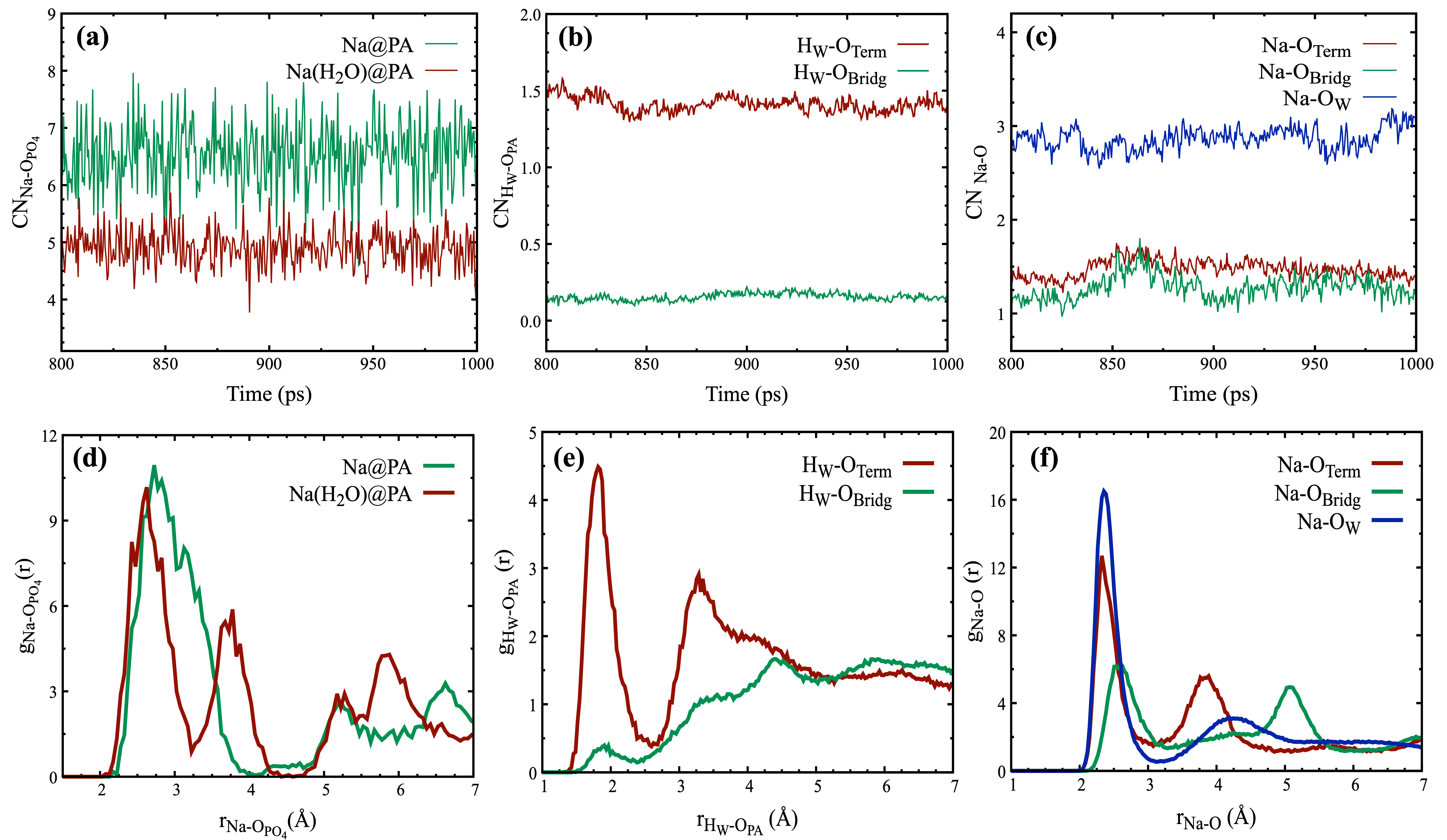}
    \caption{(Top row) Equilibrium MACE MD simulated time evolution of the CNs for (a) the enscapsulated Na$^+$-O$_{PO_4}$ interactions in Na@PA (green) and Na(H$_2$O)@PA (brown) (b) hydrogen of the free water molecules (H$_W$) to the terminal (brown) and bridging (green) oxygens of the PA in Na(H$_2$O)@PA and (c) the free Na$^+$ ions to the terminal (brown) and bridging (green) oxygens of the PA in Na(H$_2$O)@PA as well as water molecules (blue). The bottom panels (d-f) show the corresponding RDFs. Other plots for the Na@PA system are given in the Supplementary Material Fig. S7. See the text for more details.} 
    \label{fig4}
\end{figure*}
\begin{table}[htbp]
\centering
\renewcommand{\arraystretch}{1.5}
\caption{Equilibrium MACE MD calculated RDFs (R in~\AA) and average CNs for Na(H$_2$O)@PA and Na@PA systems. Interior refers to the interactions inside the PA cavity, while exterior corresponds to the interactions at the periphery of PA in aqueous solution. See the text for more details.}
\label{tab:equil}
\begin{tabular}{cc ccc cc cccc}
\toprule
\specialrule{0.1em}{0pt}{0pt}
\multirow{2}{*}{\textbf{System}} & & &\multicolumn{3}{c}{\textbf{Na(H$_2$O)@PA}}&\multicolumn{1}{c}{} & \multicolumn{3}{c}{\textbf{Na@PA}} \\
\Xcline{4-6}{1.2pt}\Xcline{9-10}{1.2pt}
 && Pair && R (\AA) & CN (Avg.) & & & R (\AA) & CN (Avg.) \\
\midrule
\specialrule{0.1em}{0pt}{0pt}
\multirow{3}{*}{\textbf{Interior}} 
   &&Na--O$_{\mathrm{PO_4}}$& &2.63 & 4.89 $\pm$ 0.32&   & & 2.73 & 6.46 $\pm$ 0.58 \\
   &&Na--O$_\mathrm{W}$& & 2.30 & 0.77 $\pm$ 0.12 & &  & - & - \\
   &&H$_\mathrm{W}$--O$_{\mathrm{PO_4}}$ & & 1.81 & 3.55 $\pm$ 0.23 & & & - & -\\
   \midrule
   \specialrule{0.1em}{0pt}{0pt}
   \multirow{5}{*}{\textbf{Exterior}}
   &&H$_\mathrm{W}$--O$_{\mathrm{Term}}$ & &1.79 & 1.41 $\pm$ 0.02 & &  & 1.82 & 0.99 $\pm$ 0.04 \\
   &&H$_\mathrm{W}$--O$_{\mathrm{Bridg}}$& &1.91 & 0.16 $\pm$ 0.05  &   &  & 1.91 & 0.09 $\pm$ 0.02 \\
   &&Na--O$_{\mathrm{Term}}$ & &2.32 & 1.48 $\pm$ 0.09 &   &   & 2.35 & 1.61 $\pm$ 0.06 \\
   &&Na--O$_{\mathrm{Bridg}}$  && 2.51 & 1.28 $\pm$ 0.13   &   &   & 2.75 & 0.99 $\pm$ 0.04 \\
   &&Na--O$_\mathrm{W}$   &&2.38 & 2.86 $\pm$ 0.12    &    &   & 2.35 & 2.38 $\pm$ 0.13 \\
\specialrule{0.1em}{0pt}{0pt}
\bottomrule
\end{tabular}
\end{table}
\section{\label{sec3:result}Results and Discussion}
To understand the ion capture and transport mechanism of the representative Na$^+$ ion in aqueous PA solutions, equilibrium and biased MD simulations were carried out using the validated MACE-MPA-0\_{medium} model, benchmarked against AIMD. Our initial analyses focus on equilibrium properties with biased MACE MD simulated results for the two considered mechanisms discussed in the following section.

\subsection{\label{sec31:mech} Equilibrium MACE MD Simulations}
The coordination environments of Na$^+$ ions in the Na@PA and Na(H$_2$O)@PA systems are investigated using average CNs and RDFs for the first solvation shells, extracted from equilibrium MACE MD simulations (see Fig. \ref{fig4} and Table \ref{tab:equil}). The nature of interactions within the interior of the PA cavity compared to its exterior is discussed separately below. 

\subsubsection{Interior Interactions}
The encapsulated Na$^+$ ion in Na(H$_2$O)@PA is coordinated to a confined water molecule on one side and interacts with an average of 4.89 $\pm$ 0.32 oxygens of the PO$_4$ templating anions from the other side (see Table \ref{tab:equil} and Fig. \ref{fig4}a). 
The calculated Na-O$_{PO_4}$ CN from MACE MD equilibrium simulations for Na(H$_2$O)@PA shows minor fluctuations over time (Fig. \ref{fig4}a). The calculated RDFs for the Na-O$_{PO_4}$ pair show two sharp peaks at $\approx$ 2.65~\AA~and $\approx$ 3.75~\AA~(Fig. \ref{fig4}d), with the first peak arising from interactions with the PO$_4$ oxygens on the same side as Na$^+$, and the second, from those on the opposite, water-facing side.

In comparison, the average CN for the same Na-O$_{PO_4}$ pair in the Na@PA system (i.e., the system without confined water) increases significantly to 6.46 $\pm$ 0.58 (Table \ref{tab:equil} and Fig. \ref{fig4}a), indicating that Na$^+$ can access more PO$_4$ oxygens inside the cavity at 298K.
The calculated Na-O$_{PO_4}$ RDF shown in Fig. \ref{fig4}d also gives a broad peak at $\approx$ 2.75 \AA~for the first solvation shell. This is simply because in Na(H$_2$O)@PA, the confined water molecule limits Na$^+$ interactions with PO$_4$ oxygens on the water-facing side, reducing its overall coordination.
Interestingly, the average CN between the encapsulated Na$^+$ and the confined water molecule in Na(H$_2$O)@PA is less than 1 (0.77 $\pm$ 0.12, Table \ref{tab:equil}). This is likely due to the competition between O$_W$ and O$_{PO_4}$ atoms for coordination to the encapsulated Na$^+$ ion. Meanwhile, hydrogens of the water molecule (H$_W$) form strong hydrogen bonds (H-bonds) with the oxygens of PO$_4$, with an average CN of 3.55 $\pm$ 0.23 and a first solvation shell distance of 1.81~\AA~(Table \ref{tab:equil} and Supplementary Material Fig. S8). These H-bond interactions stabilize the Na(H$_2$O) unit inside the cavity and shield Na$^+$ ion from forming strong interactions with the PO$_4$ templating anions from both sides of the cavity, as observed in Na@PA.

\subsubsection{Exterior Interactions}
On the exterior, PA can form H-bonds with bulk water through its terminal and bridging oxygens and/or adsorb free Na$^+$ ions. In agreement with Ref. \citenum{poblet_pa}, the H-bonds formed between water molecule and the more accessible terminal oxygens (H$_W$–O$_{Term}$) in both Na(H$_2$O)@PA and Na@PA systems are stronger than those of the bridging oxygens (see Table \ref{tab:equil}). The average CNs are slightly higher for the Na(H$_2$O)@PA than for Na@PA (1.41 $\pm$ 0.02 vs. 0.99 $\pm$ 0.04). 
Calculated RDFs also show stronger H$_W$–O$_{Term}$ H-bonds ($\approx$ 1.79–1.82~\AA) compared to H$_W$–O$_{Bridg}$ H-bonds ($\approx$ 1.91 \AA) (Table \ref{tab:equil}, Fig. \ref{fig4}e).

For Na$^+$ interactions with the terminal and bridging oxygens of the Na(H$_2$O)@PA and Na@PA systems, the average bond lengths for Na$^+$– O$_{Term}$ are found to be similar (2.32~\AA~ and 2.35~\AA), while the Na$^+$– O$_{Bridg}$ bonds are significantly shorter in Na(H$_2$O)@PA than Na@PA (2.51~\AA~vs. 2.75~\AA, Table \ref{tab:equil}). The CNs show that, like H$_W$-O$_{PA}$ interactions, Na$^+$ forms slightly stronger coordination bonds with the more accessible terminal oxygens than with bridging ones.
Both Na(H$_2$O)@PA and Na@PA systems show free Na$^+$ to O$_W$ interactions, with average CNs of 2.86 $\pm$ 0.12 to 2.38 $\pm$ 0.13 and similar Na-O$_W$ RDFs ranging from 2.38~\AA~to 2.35~\AA. 
These analyses show that with 109 water molecules present in the simulation cell, free Na$^+$ ions are rather similarly solvated by both bulk water and PA (Fig. \ref{fig4}f). Interestingly, the confined water in Na(H$_2$O)@PA is indirectly influencing PA interactions with the environment (Table \ref{tab:equil}, Fig. \ref{fig4}, and Supplementary Material Fig. S7). Considering the interactions of free Na$^+$ ions with both oxygens of water and PA combined, one arrives at CNs of 5.62 in Na(H$_2$O)@PA and 4.98 in Na@PA (Table \ref{tab:equil}). These values are comparable to the known CN of 5 to 6 for Na$^+$ ions in bulk water. Over the total time of 1ns, the encapsulated Na$^+$ ion was found to remain stable within the cavity, suggesting that the ion capture and transport is a rare event at 298K. Therefore, to further investigate this, biased MACE MD simulations were performed, the results of which are discussed below.

\subsection{\label{sec31:mech} WT-MetaD Simulations for Ion vs. Water Transport}
Using biased MACE MD simulations, two transport mechanisms are explored in this work, including the ``associative'' and ``dissociative'' pathways. In the associative mechanism, an ion exchange occurs where a free Na$^+$ ion from the aqueous solution enters the cavity from the water window, with the already encapsulated ion exiting from the opposite side. In the dissociative mechanism, the encapsulated ion exits without another ion entering. In both cases, the confined water remains inside the cavity. Structural and energetic details are discussed below. 

\subsubsection{\label{sec31:mech} Associative Mechanism for Ion Transport}
\begin{figure}[!t]
\includegraphics [width=\linewidth]{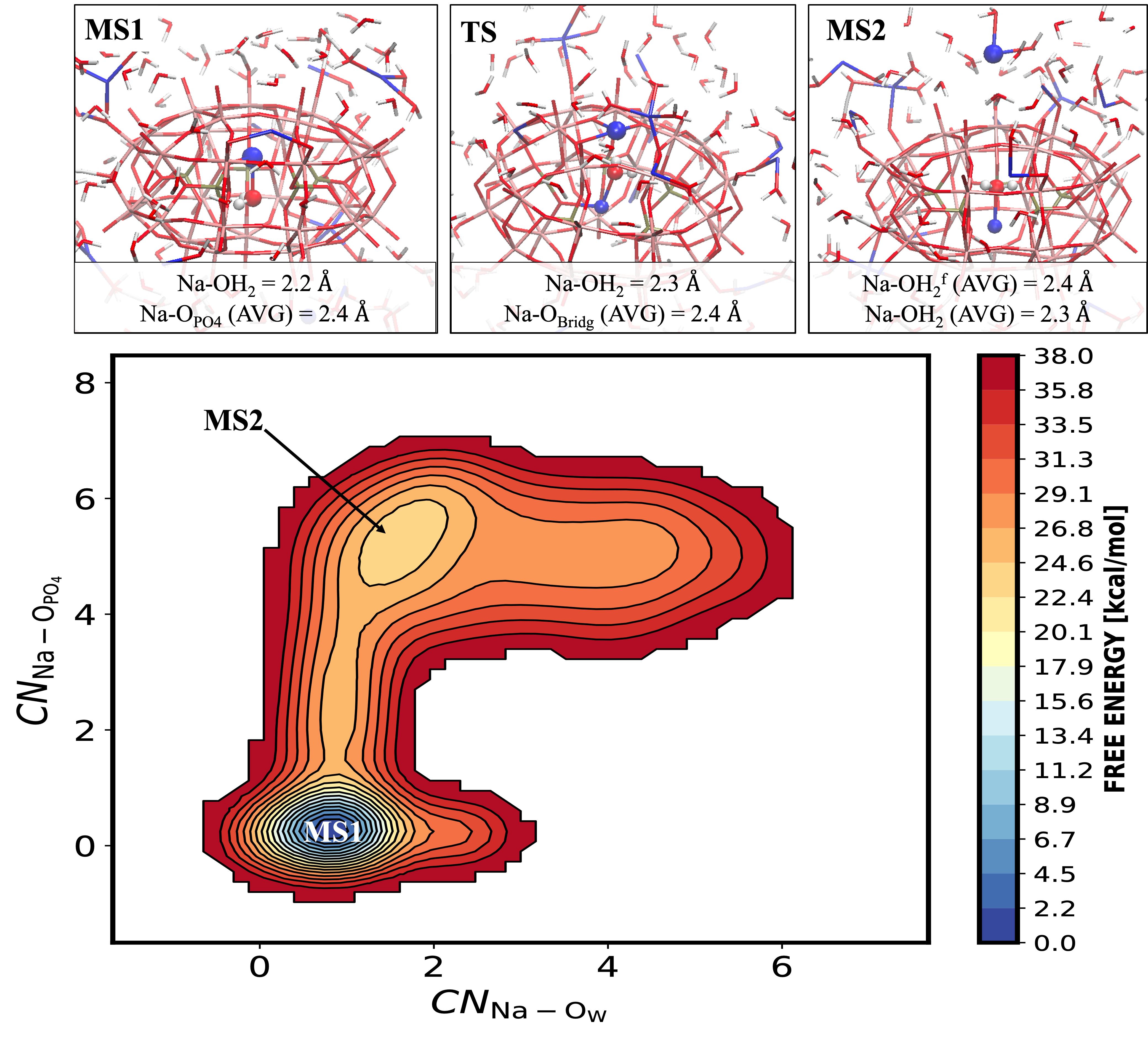}
    \caption{MACE WT-MetaD calculated FES (in kcal/mol) for ion transport through the associative mechanism. Representative snapshots for MS1, TS, and MS2 states are also given, with the main geometric parameters shown. See the text for more details.} 
    \label{fig5}
\end{figure}
\begin{table}[htbp]
\centering
\small
\renewcommand{\arraystretch}{1.5}
\caption{MACE WT-MetaD calculated changes in the free energies ($\Delta$F) and average CNs for the Na(H$_2$O)@PA and Na@PA systems considering the associative (Ass.) and dissociative (Dis.) mechanisms. For the Na$^+$ ion, both the associative (Ass. Na$^+$) and dissociative (Dis. Na$^+$) mechanisms are considered in Na(H$_2$O)@PA, while the data are tabulated for the dissociative mechanism in Na@PA. Data for ejecting confined water in Na(H$_2$O)@PA through the dissociative mechanism is also shown (Dis. H$_2$O). The superscripts and subscripts in TS show the changes (rounded to whole numbers) in CNs for CV1 and CV2, respectively. See the text for more details.}
\label{tab:wtmetad}
\begin{adjustbox}{width=\linewidth}
\begin{tabular}{cccccccccccc}
\toprule
\specialrule{0.1em}{0pt}{0pt}
\multirow{2}{*}{\textbf{System}} & & &\multicolumn{4}{c}{\textbf{Na(H$_2$O)@PA}}&\multicolumn{1}{c}{} \\
\Xcline{3-9}{1.2pt}
 &&   && $\Delta$F &CN1 & &CN2 & \\
  && &  & (kcal/mol) &(Avg.) & & (Avg.) & \\
\midrule
\specialrule{0.1em}{0pt}{0pt}
\multirow{3}{*}{\textbf{Ass. Na$^+$}}
   &&\textbf{MS1}&  & 0 & 0.92  $\pm$ 0.04  & & 0.19 $\pm$ 0.14 & \\
   &&TS$_{0\longrightarrow 6}^{1\longrightarrow 2}$& & 26.8 &0.91 $\pm$ 0.06 &  & 2.87 $\pm$ 0.83 & \\
   &&\textbf{MS2} & & 24.6 &1.87 $\pm$ 0.22  & & 5.30 $\pm$ 0.46&\\
   \midrule
   \specialrule{0.1em}{0pt}{0pt}
   \multirow{3}{*}{\textbf{Dis. Na$^+$}}
   &&\textbf{MS1}& & 0 & 4.89 $\pm$ 0.55  & & 0.77 $\pm$ 0.29 & \\
   &&TS$_{1\longrightarrow 2}^{5\longrightarrow 0}$& & 21.3 &0.47 $\pm$ 0.13 &  & 1.29 $\pm$ 0.01 & \\
   &&\textbf{MS2} &  & 19.2 &0.38 $\pm$ 0.07  & & 1.89 $\pm$ 0.30&  \\
   \midrule
\specialrule{0.1em}{0pt}{0pt}
\multirow{3}{*}{\textbf{Dis. H$_2$O}}
   &&\textbf{MS1}& & 0 & 4.62 $\pm$ 0.44  & & 2.00 $\pm$ 0.00 & \\
   &&TS$_{2\longrightarrow 6}^{5\longrightarrow 0}$&  & 25.6 &1.21 $\pm$ 0.20 &  & 2.18 $\pm$ 0.09 &  \\
   &&\textbf{MS2} &  & 19.2 &0.01 $\pm$ 0.01 &&4.40 $\pm$ 1.01&  \\
   \midrule
   \specialrule{0.1em}{0pt}{0pt}
\multirow{2}{*}{\textbf{System}} & & &\multicolumn{4}{c}{\textbf{Na@PA}}&\multicolumn{1}{c}{} &\\
\Xcline{3-9}{1.2pt}
   \multirow{3}{*}{\textbf{Dis. Na$^+$}}
   & &\textbf{MS1}& & 0 & 5.76 $\pm$ 0.46  & & 0.02 $\pm$ 0.00 \\
   & &TS$_{0\longrightarrow 3}^{6\longrightarrow 0}$ &  &23.5&  1.30 $\pm$ 0.49  & & 0.35 $\pm$ 0.20 \\
 & &\textbf{MS2} & & 17.1 & 0.05 $\pm$ 0.05  & & 2.79 $\pm$ 0.57 \\
 \specialrule{0.1em}{0pt}{0pt}
\bottomrule
\end{tabular}
\end{adjustbox}
\end{table}
In the initial configuration for the first metastable state (MS1), Na$^+$ ion in Na(H$_2$O)@PA resides in the side cavity of PA and is coordinated to the confined H$_2$O molecule as illustrated at the top of Fig. \ref{fig5}. 
Equilibrium MD analysis showed strong electrostatic interactions between Na$^+$ and oxygens of PO$_4$ (O$_{PO4}$) on one side, while the confined water forms hydrogen bonds with O$_{PO_4}$ from the other side. As mentioned, in the associative mechanism, a Na$^+$ ion from solution is encapsulated, followed by the ejection of the already encapsulated one. As mentioned, this ion exchange process is simulated using two CVs: (i) CN between the encapsulated Na$^+$ ion and the surrounding water oxygens CN$_{Na-O_W}$, and (ii) CN between the free Na$^+$ ions and O$_{PO_4}$, i.e., CN$_{Na^f-O_{PO_4}}$. 

A detailed analysis of the trajectory reveals that, as the reaction proceeds, exterior Na$^+$ ions approach the cavity by first interacting with the terminal oxygens of PA, while the initially encapsulated Na$^+$ remains stable inside. One of the Na$^+$ ions eventually enters through the water-facing window, forming a solvent-separated ion pair (SSIP) like intermediate species, where both Na$^+$ ions strongly interact with the confined H$_2$O (see Fig. \ref{fig5}). The initially encapsulated Na$^+$ ion is then ejected into the solution. 
The calculated FES shows the lowest energy basin at MS1, where the initial Na$^+$ is solely coordinated to the confined water, with CN$_{Na-O_W}$ = 0.92 $\pm$ 0.04 and CN$_{Na^f-O_{PO_4}}$ = 0.19 $\pm$ 0.14 (see Fig. \ref{fig5} and Table \ref{tab:wtmetad}).

The transition state (TS) is characterized by CN$_{Na-O_W}$ fluctuating around 1 and CN$_{Na^f-O_{PO_4}}$ increasing to $\approx$ 2-3, with average calculated CNs of 0.91 $\pm$ 0.06 and 2.87 $\pm$ 0.83 (see Table \ref{tab:wtmetad}). 
This indicates that the original Na$^+$ remains inside the cavity while a free Na$^+$ enters and coordinates with $\approx$3 oxygen atoms of the PO$_4$ units. 
The free energy barrier for this considered associative mechanism is calculated as 26.8 kcal/mol. After passing through the barrier, the system reaches the second metastable state (MS2) at 24.6 kcal/mol, where the encapsulated Na$^+$ ion has fully exited the cavity. In MS2, CN$_{Na-O_W}$ is $\approx$2, and CN$_{Na^f-O_{PO_4}}$ is $\approx$5–6, with average values of 1.87 $\pm$ 0.22 and 5.30 $\pm$ 0.46, respectively. This confirms that the pre-encapsulated Na$^+$ ion exits the cavity and coordinates to the exterior water molecules while the newly encapsulated Na$^+$ ion is now strongly bonded to O$_{PO_4}$. These calculated CNs are close to those from equilibrium MACE MD simulations reported in Table \ref{tab:equil}.

This rather high calculated energy barrier raises an important question: What would be the free energy barrier of ejecting a Na$^+$ ion from the cavity in the absence of another incoming ion? In other words, can one eject the ion by simply heating the PA solution in deionized water? This question motivated us to further investigate an alternative ``dissociative'' pathway focusing on the structural and free energy changes involved in releasing the encapsulated Na$^+$ ion directly into the solution.

\subsubsection{\label{sec31:mech} Dissociative Mechanism for Ion Transport}
\begin{figure*}[!t]
\includegraphics [width=\textwidth]{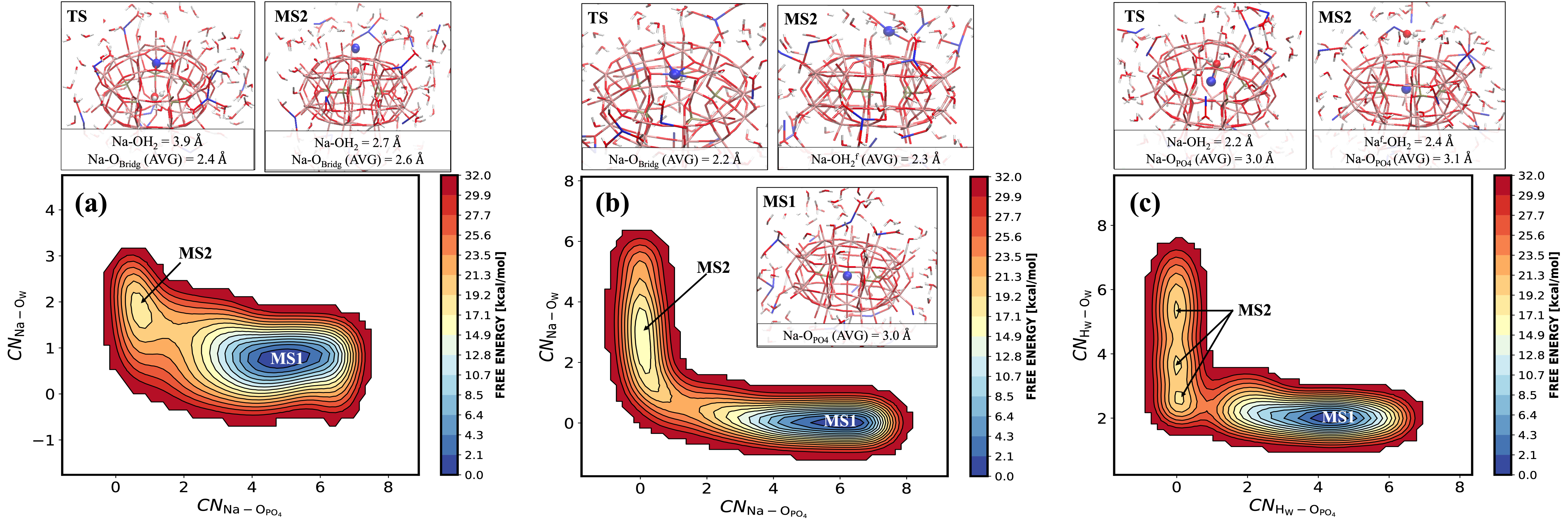}
    \caption{Calculated FESs (in kcal/mol) for the dissociative mechanism involving Na$^+$ ion in (a) Na(H$_2$O)@PA and (b) Na@PA systems. Panel (c) shows the calculated FES for ejecting the confined water in Na(H$_2$O)@PA. Representative snapshots for MS1, TS, and MS2 states are also given, with the main geometric parameters shown. See the text for more details.}
    \label{fig6}
\end{figure*}

The dissociative transport mechanisms were examined by defining two new CVs based on the CN of the encapsulated Na$^+$ ion to (i) O$_{PO_4}$ (i.e., CN$_{Na-O_{PO_4}}$ and (ii) oxygens of bulk water (O$_W$) (i.e., CN$_{Na-O_W}$). The simulated FES for the Na(H$_2$O)@PA system is shown in Fig. \ref{fig6}a. The initial MS1 configuration is the same as that of the associative pathway. In the TS, the confined $Na^+$ begins to detach from the water while concurrently interacting with both O$_{PO_4}$ and the bridging oxygens ($O_{bridg}$) of PA.
In the TS, the calculated CN$_{Na-O_{PO_4})}$ drops toward zero, while CN$_{Na-O_{W}}$ rises to about 2. The average values, 0.47 $\pm$ 0.13 and 1.29 $\pm$ 0.01, indicate that the encapsulated Na$^+$ is detaching from the O$_{PO_4}$ units and beginning to interact with bulk water (see Table \ref{tab:wtmetad}). The free energy barrier from MS1 to TS for this dissociative mechanism is estimated to be $\approx$21.3 kcal/mol. This value matches that of the free energy barrier obtained from reference AIMD simulations (see Supplementary Material Fig. S3).
As expected, in MS2, with a calculated $\Delta$F of 19.2 kcal/mol, CN$_{Na-O_{PO_4}}$ drops to zero while CN$_{Na-O_W}$ rises to $\approx$ 2. The average values, 0.38 $\pm$ 0.07 and 1.89 $\pm$ 0.30-confirm the complete ejection of the Na$^+$ into the solution, leaving the water molecule behind (Fig. \ref{fig6}a).\\

\begin{figure*}[!t]
\includegraphics [width=\linewidth]{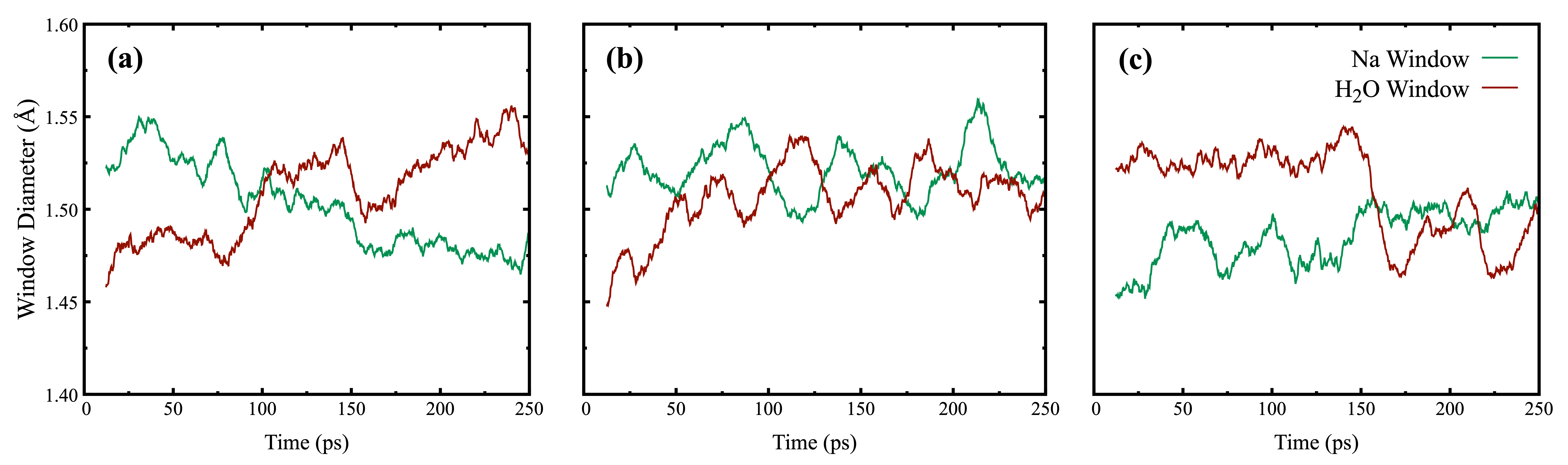}
    \caption{MACE MD calculated time evolution of the PA window diameters during the considered (a) associative mechanism for Na(H$_2$O)@PA, (b) dissociative mechanism for Na$^+$ in Na(H$_2$O)@PA, and (c) dissociative mechanism for ejecting H$_2$O in Na(H$_2$O)@PA.} 
    \label{fig7}
\end{figure*}

This calculated barrier is 5.5 kcal/mol lower than that of the associative mechanism discussed in the previous section. Analysis of the window diameters averaged across trajectories from all five walkers using the pywindow code\cite{miklitz2018pywindow} shows that the Na (H$_2$O) window contracts (expands) by $\approx$ 0.1~\text{\AA} during ion entry and exit (see Fig. \ref{fig7}a). In contrast, during the dissociative mechanism for the Na$^+$ ion, the changes in the diameters of both windows are $\approx$ 0.05~\AA~which is about half (see Fig. \ref{fig6}b). We recognize that the implication of these results for ion capture using PA is rather significant as it shows efficient ion capture and release by PA should be performed in two separate steps: (i) heating in deionized water to eject the already encapsulated ion followed by (ii) transferring the POM adsorbent to the target solution for the selective ion capture. We, however, caution against generalizing these results to other POMs with different windows and cavity sizes.

\subsubsection{Effects of the Presence of the Confined Water} 
As shown from our equilibrium MACE MD data, water coordination to the encapsulated ion inside the cavity has a significant impact on its interactions within PA. As such, water is expected to affect the calculated FESs for ion capture and transport in the Na(H$_2$O)@PA system. As mentioned previously, in Na-H$_2$O@PA, the encapsulated Na$^+$ ion is stabilized by interactions with both confined water and O$_{PO_4}$, with an average CN$_{Na-O_{PO_4}}$ of 4.89 $\pm$ 0.32 (see Table \ref{tab:equil}). In contrast, in Na@PA (i.e., without confined water), the coordination increases to 6.46 $\pm$ 0.58.
To assess the impact of confined water, the FES was calculated for the dissociative mechanism in Na@PA using the same two CVs as for Na(H$_2$O)@PA (Fig. \ref{fig6}b). In MS1, CN$_{Na-O_{PO_4}}$ and CN$_{Na-O_W}$ are calculated to be 5.76 $\pm$ 0.46 and 0.02 $\pm$ 0.00, respectively (Table \ref{tab:wtmetad}). At the TS, these change to 1.30 $\pm$ 0.49 and 0.35 $\pm$ 0.20, indicating partial ejection of the Na$^+$ ion from the cavity, though it still remains near the window and has not yet fully ejected into the solution.

The calculated free energy barrier in Na@PA is estimated to be $\approx$23.5 kcal/mol. In MS2 (17.1 kcal/mol), as expected, CN$_{Na-O_{PO_4}}$ drops to near zero (0.05 $\pm$ 0.05), while CN$_{Na-O_W}$ increases to 2.79 $\pm$ 0.57, confirming complete ion release into the aqueous solution. This barrier is around $\approx$2.2 kcal/mol higher than the Na(H$_2$O)@PA system. This can likely be attributed to water shielding Na$^+$ from interacting with some of the O$_{PO_4}$ sites, as seen from our equilibrium MACE MD data. Our RDF analyses (Fig. (\ref{fig4}d)) further showed that the confined water restricts ion mobility, pushing Na$^+$ toward a side pocket in Na-H$_2$O@PA.
As mentioned, the calculated RDFs for Na–H$2$O@PA reveal two distinct Na–O$_{PO_4}$ peaks at $\approx$2.75~\AA~ and $\approx$3.75~\AA, showing asymmetric coordination due to water-induced ion displacement. In contrast, Na@PA showed a single broad peak, indicating a delocalized Na$^+$ interacting with almost all 10 O$_{PO_4}$ sites. As such,
the confined water's role is to weaken Na$^+$–O$_{PO_4}$ interactions, reducing its CN and facilitating ion release, as confirmed by both equilibrium and biased MD data. The ion must also break its bond with the confined water in order to freely diffuse into the solution. The lower energy cost of breaking a single Na-H$_2$O bond (CN$_{Na-O_W}$ = 0.77 $\pm$ 0.12) compared to the drop in Na–O$_{PO_4}$ coordination from 6.46 $\pm$ 0.58 to 4.89 $\pm$ 0.32 (see Table \ref{tab:equil}) helps offset this, explaining the small energy barrier difference. 
Overall, confined water is found to enhance ion mobility through lowering the free energy barrier to ion capture and release in PA. The role of water in shielding the ion from the oxygens of POMs is likely similar in other POM clusters and their extended frameworks and should be explicitly taken into account when studying transport in these intriguing systems.

\subsubsection{Dissociative Mechanism for Water Transport}

As shown earlier, confined water weakens Na$^+$–O$_{PO_4}$ interactions, lowering the free energy barrier by 2.2 kcal/mol. These prompted us to explore the FES for ejecting confined water itself. Two CVs were selected, based on the CN of the hydrogen atoms of the confined water (H$_W$) to: (i) O$_{PO_4}$ (i.e., CN$_{H_W-O_{PO_4}}$), and (ii) oxygens of the bulk water (i.e., CN$_{H_W-O_W}$).
In MS1, the confined water strongly interacts with the encapsulated Na$^+$ ion and forms strong H-bonds with O$_{PO_4}$ with an average CN of 4.62 $\pm$ 0.44 (Table \ref{tab:wtmetad}), stabilizing itself within the cavity. 
At the TS, CN$_{H_W-O_{PO_4}}$ drops to 1.21 $\pm$ 0.20 while CN$_{H_W-O_W}$ rises to 2.18 $\pm$ 0.09, indicating partial dissociation from the oxygens of the templating anions. The free energy barrier for water ejection is $\approx$ 25.6 kcal/mol, which is 4.3 kcal/mol higher than that for Na$^+$ in Na(H$_2$O)@PA (see Table \ref{tab:wtmetad} and Fig. \ref{fig6}c).

In MS2 at 19.2 kcal/mol, CN$_{H_W-O_{PO_4}}$ drops to near zero, while CN$_{H_W-O_W}$ rises to 4.40 $\pm$ 1.01. Interestingly, MS2 reveals three distinct minima with calculated CN$_{H_W-O_W}$ values of $\approx$ 2 (still water is interacting with PA bridging oxygens), $\approx$ 3 (transitioning to out of the window), and $\approx$ 5 (full solvation in bulk aqueous environment). The calculated 2.1 kcal/mol energy barrier that separates these three minima shows that all of them should be accessible at room temperature.
The higher calculated barrier of 4.3 kcal/mol for water ejection likely arises from its larger kinetic diameter ($\approx$2.65) compared to bare Na$^+$ ($\approx$2.02~\AA), making transport through the PA window more difficult. Fig. \ref{fig7}c shows that, when ejecting water, the change in the calculated water window averaged over all five walkers is $\approx$ 0.1~\AA, with the Na window staying more or less the same. This shows that upon heating Na(H$_2$O)@PA system in deionized water, Na$^+$ will likely be the first species that is going to be released into the solution before confined water can be ejected. 

\section{\label{sec4:conc}Conclusions}
Using equilibrium and biased MACE MD simulations, carefully validated against reference AIMD data, we have studied the aqueous solution chemistry of the ion capture and transport mechanisms in PA as the smallest member of the extended POM family with a cavity large enough to host transition metal ions. Comparisons between Na(H$_2$O)@PA to Na@PA systems in aqueous solutions showed that confined water has a significant effect by shielding the encapsulated Na$^+$ ion from interactions with oxygens of the PO$_4$ templating anions. Through well-converged multiple-walker WT-MetaD simulations, two different associative and dissociative ion transport mechanisms were carefully investigated. Our detailed analyses showed that the ion capture in PA occurs through a dissociative mechanism with a free energy barrier of 21.3 kcal/mol. 
Results obtained from this work demonstrate the critical role of internal hydration within the PA cavity in tuning ion exchange and transport processes, offering important insights into the dynamic behaviour of ion capture and release in POMs. Our future experimental and theoretical studies will focus on exploring how encapsulated hydronium affects ion exchange and also extending these studies to 2D and 3D frameworks of PA and similar POMs.
%Results obtained from this study are helpful in providing a molecular-level insights into the water H-bond network and how it affects the ion transport mechanisms in POMs with internal cavities including PA and the larger P$_8$W$_{48}$ POMs.
\section*{Supplementary Material}
Details of structural parameters from geometry optimizations from AIMD, radial distribution functions (RDFs) from both AIMD and MACE MD for benchmarking and production runs, time evolution of coordination numbers from equilibrium MD simulations for Na@PA, schematic representations of collective variables (CVs), switching functions, and cutoff radius ($r_c$) fittings, as well as the time evolution of CVs from MACE MD for all considered systems, and the free energy surface (FES) plot calculated from reference AIMD along with its corresponding time evolution of CVs.

\begin{acknowledgments}
This work was supported by the U.S. Department of Energy Office of Science, Basic Energy Sciences Program (Grant No. DE-SC0024512). PKT thanks DOE Nuclear Energy – Materials Research and Waste Form Development Campaign. Particularly, PKT thanks Dr. Ken Marsden (INL), Mrs. Amy Welty (INL), and Mrs. Kimberly Gray (DOE Nuclear Energy). ZP and MRM thank UMKC Tier 1 Funding For Excellence (FFE) grant for support. Simulations presented in this work used resources from Bridges-2 at Pittsburgh Supercomputing Center through allocation PHY230099 and PHY240029P from the Advanced Cyberinfrastructure Coordination Ecosystem: Services \& Support (ACCESS) program,\cite{access} which is supported by National Science Foundation grants \#2138259, \#2138286, \#2138307, \#2137603, and \#2138296. Technical support and computing resources provided by the HPC center at UMKC are also gratefully acknowledged.
\end{acknowledgments}

\subsection*{Conflict of Interest}

\noindent The authors have no conflicts to disclose.

\subsection*{Author Contributions}
\noindent All authors have given approval to the final version of this manuscript. 

\section*{\label{sec_contrib} DATA AVAILABILITY}
\noindent The data that support the findings of this study is provided in the main text and the accompanied Supplementary Material. Additional data are available from the corresponding author upon reasonable request.

\section*{REFERENCES}
\bibliography{bib}

%\begin{center}
%    \includegraphics[width=0.99\linewidth]{figs/toc.png}
%    TOC entry
%\end{center}

\end{document}

\newpage
\tableofcontents
\newpage

\begin{table}
\caption{Calculated W--O distances and \% errors for the optimized Na(H$_2$O)@PA system at the rev-PBE-D3 level with different CUTOFF values and with and without U$_{eff}$ compared to the experiment.}
\centering
\addcontentsline{toc}{table}{Table \ref{table:simple}. Geometric parameters and \% errors compared to the experiment.}
\centering
\resizebox{0.99\linewidth}{!}{
%\begin{ruledtabular}
\begin{tabular}{ccccccccc}
\hline
CUTOFF & &W-O$_{Term}$ & W-O$_{Bridg} $  & W-O$_{PO_4}$  & W-O$_{Term}  $ & W-O$_{Bridg} $ & W-O$_{PO_4}$ \\ 
(Ry) & & (\AA) & (\AA) & (\AA) & (\% error) & (\% error) & (\% error)\\
\hline
&Exp. (Ref. \citenum{kim1999slow}) &1.70-1.74 & 1.80-2.09 & 2.18-2.31 \\
500 & Na(H$_2$O)@PA (no U$_{eff})$ &1.74-1.75 & 1.81-2.14 & 2.14-2.32 & 1.45 & 1.80 & 0.66  \\
500 & Na(H$_2$O)@PA (U$_{eff}$, W=1.212, O=8.794) &1.74-1.75 & 1.82-2.09 & 2.14-2.30 & 1.45 & 0.51 & 1.33  \\
500 & Na(H$_2$O)@PA (U$_{eff}$, W=1.421, O=8.794) &1.74-1.76 & 1.82-2.09 & 2.13-2.29 & 1.74 & 0.51 & 1.55  \\
550 & Na(H$_2$O)@PA (no U$_{eff})$ &1.74-1.76 & 1.81-2.16 & 2.15-2.30 & 1.74 & 2.06 & 0.89  \\
550 & Na(H$_2$O)@PA (U$_{eff}$, W=1.212, O=8.794) &1.74-1.76 & 1.82-2.12 & 2.14-2.30 & 1.74 & 1.29 & 1.11  \\
550 & Na(H$_2$O)@PA (U$_{eff}$, W=1.421, O=8.794) &1.74-1.76 & 1.83-2.11 & 2.14-2.30 & 1.74 & 1.29 & 1.11  \\
600 & Na(H$_2$O)@PA (no U$_{eff})$ &1.74-1.75 & 1.81-2.16 & 2.14-2.31& 1.45 & 2.06 & 0.89  \\
600 & Na(H$_2$O)@PA (U$_{eff}$, W=1.212, O=8.794) &1.74-1.76 & 1.82-2.13 & 2.14-2.30 & 1.74 & 1.54 & 1.11  \\
600 & Na(H$_2$O)@PA (U$_{eff}$, W=1.421, O=8.794) &1.74-1.76 & 1.84-2.09 & 2.13-2.30 & 1.74 & 1.03 & 1.34  \\
650 & Na(H$_2$O)@PA (no U$_{eff})$ &1.74-1.76 & 1.81-2.17 & 2.16-2.32& 1.74 & 2.31 & 0.22  \\
700 & Na(H$_2$O)@PA (no U$_{eff})$ &1.74-1.76 & 1.81-2.14 & 2.16-2.31& 1.74 & 1.54 & 0.44  \\
\hline
\end{tabular}}
\label{table:simple}
\end{table}

\clearpage
\noindent \textbf{Section S1. Benchmarking AIMD Simulations}
\addcontentsline{toc}{section}{Section S1. Benchmarking AIMD Simulations}
Using the rev-PBE-D3 optimized structure of Na(H$_2$O)@PA, we performed a series of 10 ps AIMD simulations using different energy cutoffs of 300Ry, 400Ry, and 500Ry. For all considered energy cutoff values, two simulation cells containing 110 and 150 H$_2$O molecules were considered. Our AIMD calculated RDFs at 298K with different energy cutoffs are found to agree very well with one another (see Figs. S1 and S2).
Upon close inspection of the equilibrated systems, the first two W-O$_{PA}$ peaks at $\approx$1.7~\AA~and $\approx$ 1.9~\AA~were assigned to W--O$_{Term}$ (the terminal oxygens of PA) and the W--O$_{Bridg}$ (the bridging oxygens of PA), respectively. The relatively weaker W--O$_{PO_4}$ (the oxygen of the PO$_4$ templating anions) peak appears at longer distances of $\approx$ 2.45~\AA, as a shoulder.
Fig. S1b shows the calculated RDFs for Na--O$_{PA}$ pairs, where again a close agreement in terms of the position of the first peak was observed. The same agreement was observed for the calculated intermolecular Na--O$_W$ (Fig. S1c) and O$_{PA}$--H$_W$ (Fig. S1g) pairs. 
The O$_W$--O$_W$ (oxygen of water), O$_W$--H$_W$ (hydrogen of water), and H$_W$--H$_W$ interactions were also closely matched across different energy cutoffs for both 110H$_2$O and the larger 150H$_2$O water systems. Given the cost of these simulations and the insensitive nature of the interactions to the choice of cutoff energy and size of the simulation cell, we selected the 300RY cutoff and the smaller 110 H$_2$O system as the best compromise in this work.

\clearpage
\begin{figure*}[!htb]
\addcontentsline{toc}{figure}{Figure \ref{fig:1}. Calculated RDFs from the AIMD simulations of the 110 H$_2$O system}
\centering
    \includegraphics [width=0.99\textwidth]{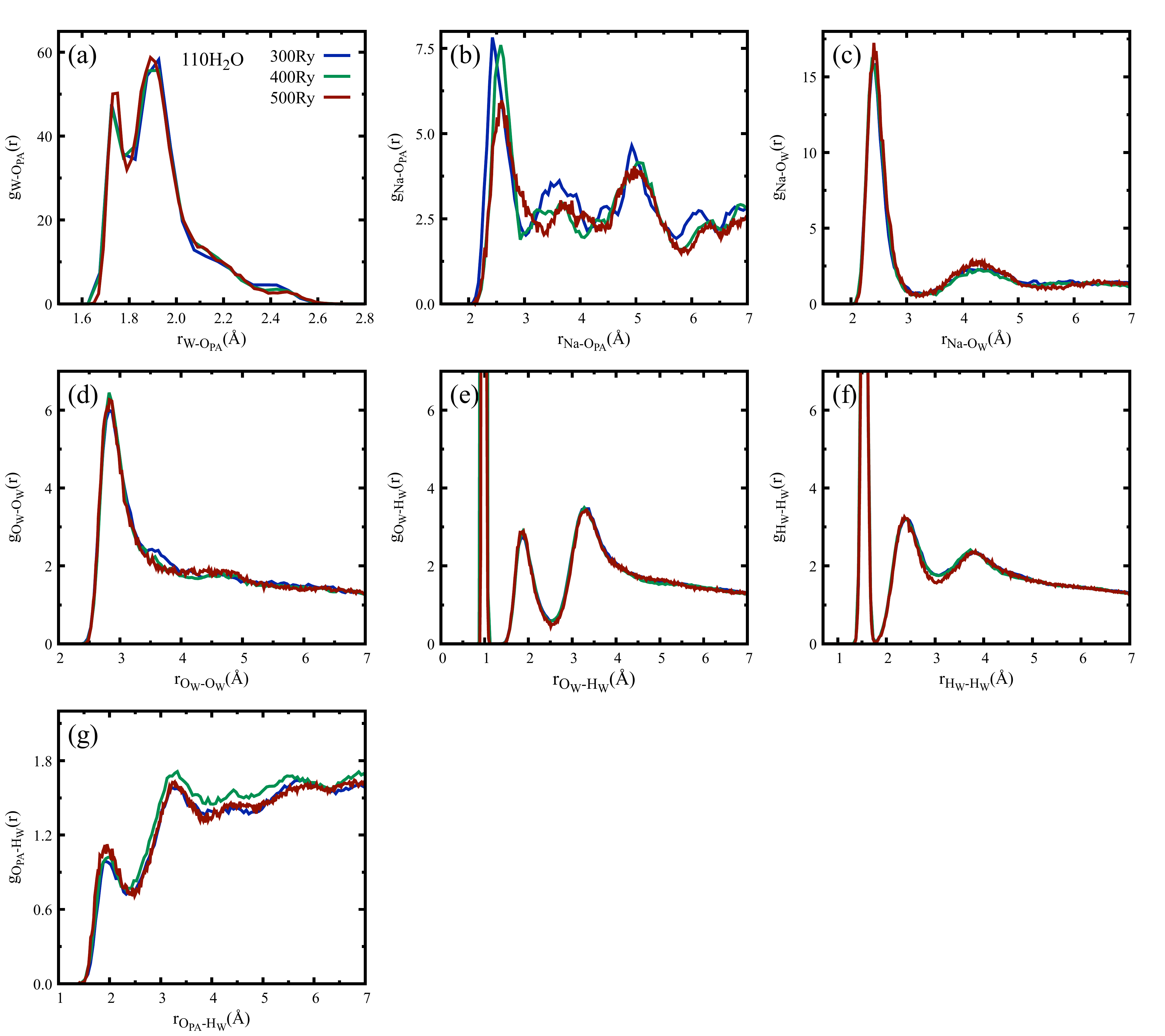}
     \caption{ Calculated RDFs for (a) W--O$_{PA}$ (oxygen of PA), (b) Na--O$_{PA}$, (c) Na--O$_W$ (oxygen of water), (d) O$_W$--O$_W$, (e) O$_W$--H$_W$ (hydrogen of water), (f) H$_W$--H$_W$, and (g) O$_{PA}$--H$_W$ pairs from NVT equilibrated 110 H$_2$O system with different energy cutoffs of 300Ry, 400Ry, and 500Ry. }   
    \label{fig:1}
\end{figure*}

\clearpage
\begin{figure*}[!htb]
\addcontentsline{toc}{figure}{Figure \ref{fig:2}. Calculated RDFs from the AIMD simulations of the 150 H$_2$O system}
\centering
    \includegraphics [width=0.99\textwidth]{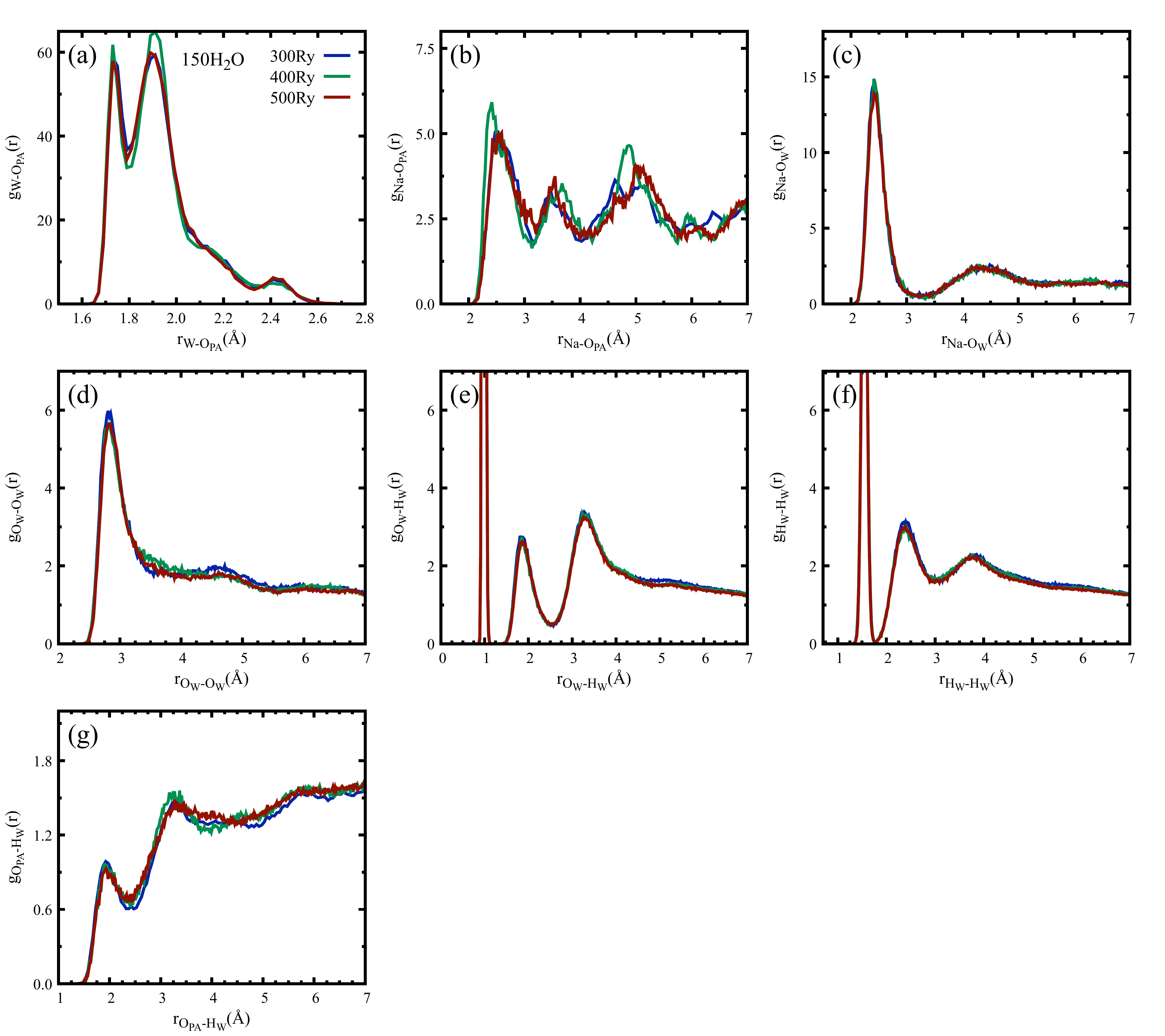}
     \caption{ Calculated RDFs for (a) W--O$_{PA}$ (oxygen of PA), (b) Na--O$_{PA}$, (c) Na-O$_W$ (oxygen of water), (d) O$_W$--O$_W$, (e) O$_W$--H$_W$ (hydrogen of water), (f) H$_W$--H$_W$, and (g) O$_{PA}$--H$_W$ pairs from NVT equilibrated 150 H$_2O$ system with different energy cutoffs of 300Ry, 400Ry, and 500Ry. }   
    \label{fig:2}
\end{figure*}

\clearpage
\begin{figure*}[!htb]
\addcontentsline{toc}{figure}{Figure \ref{fig:3}. Reference AIMD simulated FES for the dissociative mechanism}
\centering
   \includegraphics[width=0.99\textwidth,keepaspectratio]{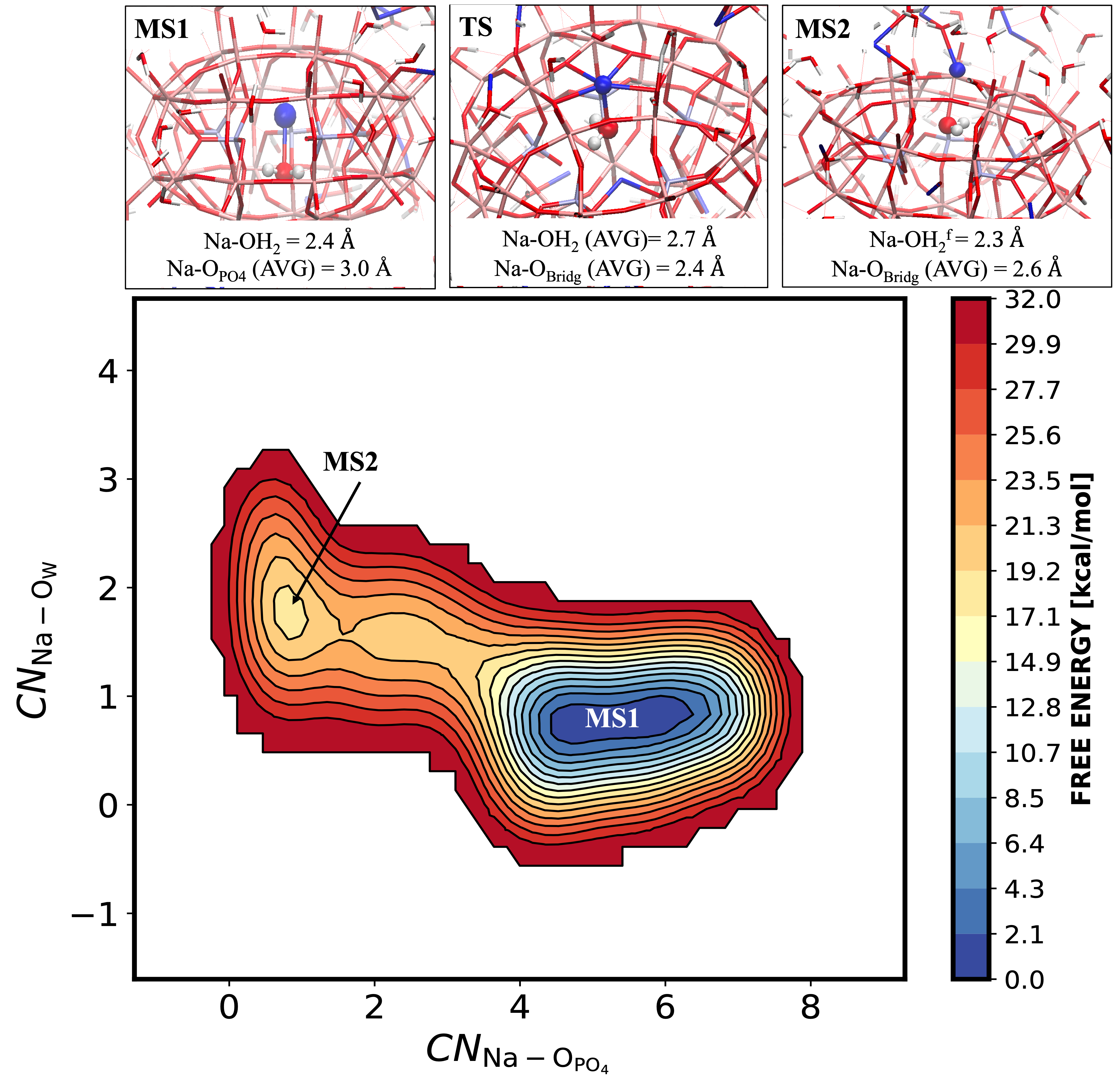}
        \caption{The rev-PBE-D3 AIMD calculated FES for ejecting the encapsulated Na$^+$ ion in Na(H$_2$O)@PA through the dissociative mechanism. 5 walkers are used in these WT-MetaD simulations, with a time of 15 ps each. MS1, TS, and MS2 snapshots are shown with the main geometric parameters given. See the main text for more details. }    
        \label{fig:3}
\end{figure*}

\clearpage
\begin{figure*}[!htb]
\addcontentsline{toc}{figure}{Figure \ref{fig:4}. Time evolution of the 5 walkers from AIMD WT-MetaD simulations}
\centering
   \includegraphics[width=0.99\textwidth,keepaspectratio]{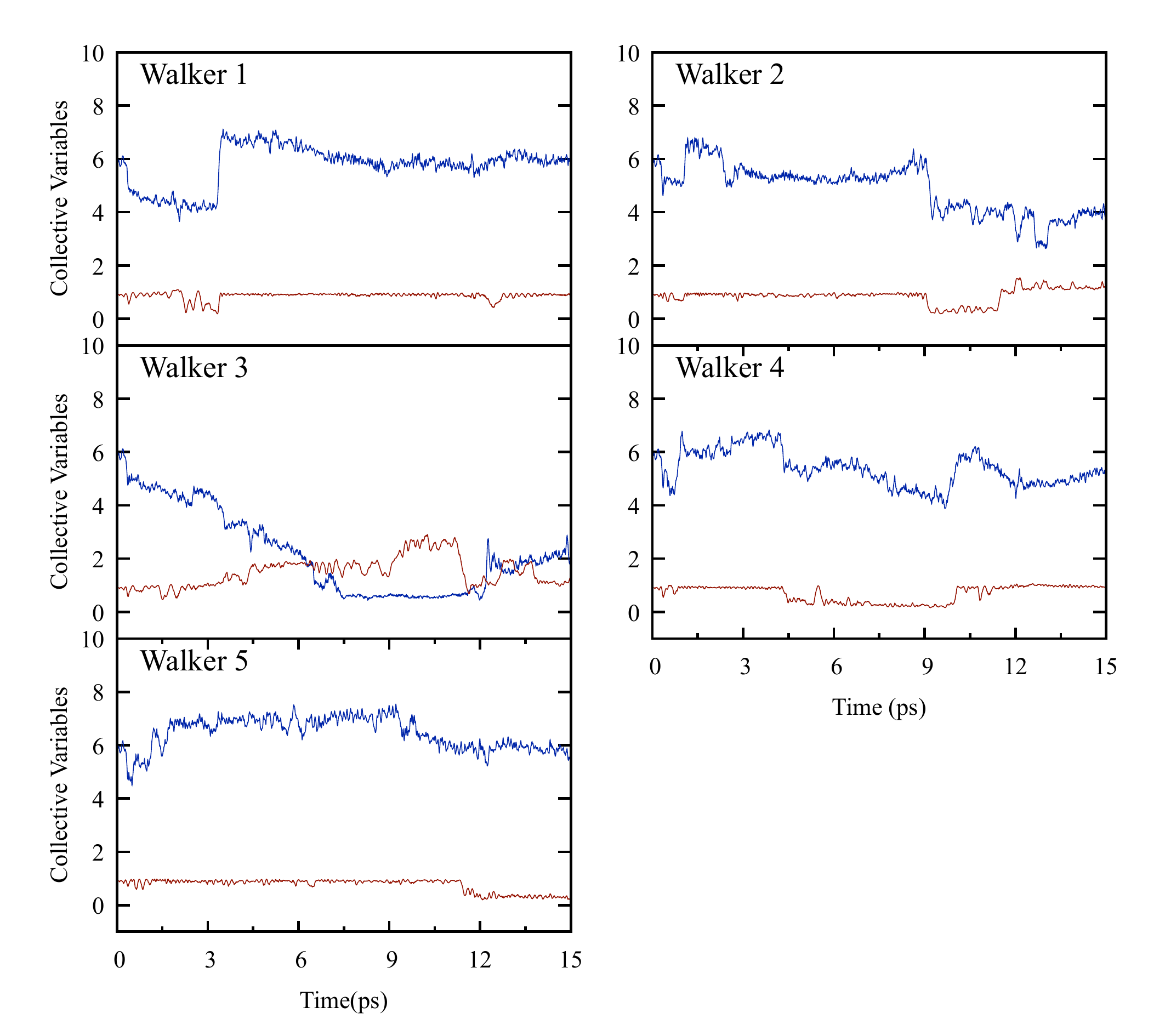}
        \caption{Time evolution of the CV1 (blue) and CV2 (brown) from the 5 walkers WT-MetaD AIMD simulations illustrated in Fig. \ref{fig:3} above. See Fig. \ref{fig:9} for the definitions of the CVs.}    
        \label{fig:4}
\end{figure*}

\clearpage
\noindent \textbf{Section S2. MACE Foundation Models}
\addcontentsline{toc}{section}{Section S2. MACE Foundation Models}
MACE is a general-purpose machine learning framework designed for developing interatomic potentials that can be applied outside the box to a diverse set of problems.\cite{batatia2022mace} It was developed by integrating the atomic cluster expansion (ACE)\cite{drautz2019atomic} approach with equivariant message passing neural networks (MPNNs). The MACE architecture, therefore, represents a state-of-the-art approach in the field of molecular dynamics, utilizing higher-order equivariant MPNNs with a detailed high-body representation of the local atomic environment through spherical harmonic polynomials to achieve fast and accurate force fields.\cite{batatia2022mace} Recently, pre-trained MACE foundation models have been introduced\cite{MACE-MP} that are capable of predicting materials properties based on their elemental compositions with comparable density functional theory (DFT) accuracies, at the Perdew-Burke-Ernzerhof (PBE) level with the Hubbard U correction. 

\begin{figure*}[!htb]
\addcontentsline{toc}{figure}{Figure \ref{fig:5}. Calculated RDFs from MACE MD compared to reference AIMD}
\centering
   \includegraphics[width=0.99\textwidth,keepaspectratio]{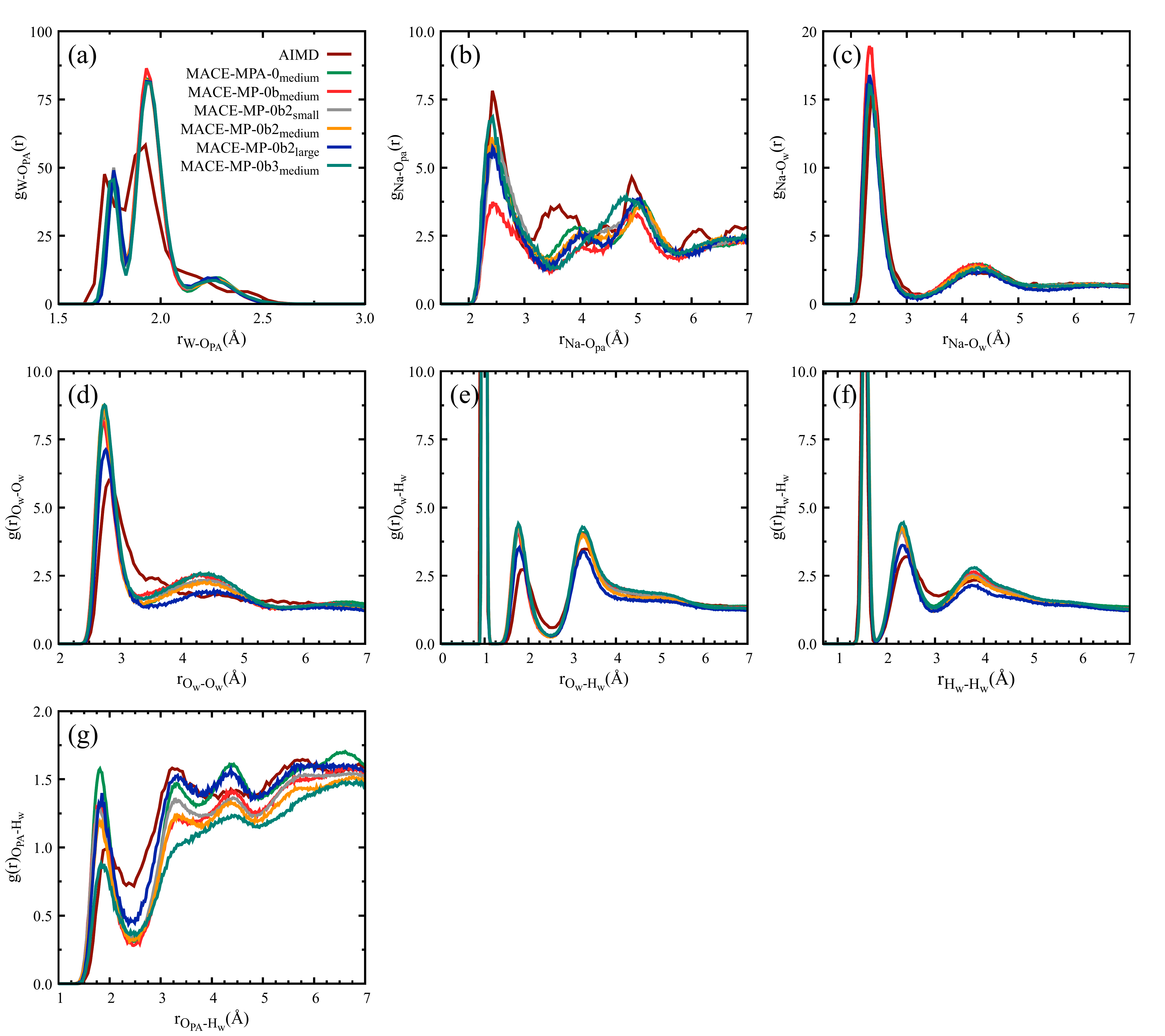}
        \caption{ Calculated RDFs for (a) W--O$_{PA}$ (oxygen of PA), (b) Na--O$_{PA}$, (c) Na--O$_W$ (oxygen of water), (d) O$_W$--O$_W$, (e) O$_W$--H$_W$ (hydrogen of water), (f) H$_W$--H$_W$, and (g) O$_{PA}$--H$_W$ pairs from 200 ps NVT equilibrium simulations for the Na(H$_2$O)@PA system with 110 H$_2$O molecules using different MACE models compared to reference AIMD.} 
        \label{fig:5}
\end{figure*}

The first-generation MACE-MP-0 models (small, medium, large) provide broad, general-purpose predictions for 89 elements but may lack accuracy in capturing certain interatomic interactions\cite{batatia2023foundation}. Each size of the model, small, medium, and large, is different in terms of cutoff radius 4.5-5.0~\AA~and maximal message equivariance (L = 0, 1, and 2). Subsequent improvements in MACE-MP-0b (small and medium) focused on enhancing pair repulsion and correcting the behavior of isolated atoms, thereby improving accuracy for materials involving strong atomic repulsion and non-interacting atoms.\cite{batatia2023foundation} The MACE-MP-0b2 version (small, medium, large) further refined the model with better treatment of interatomic interactions, offering model sizes that provide trade-offs between speed and accuracy.\cite{batatia2023foundation} The MACE-MP-0b3 (medium) version continued these improvements with a focus on maintaining stability in molecular dynamics simulations.\cite{batatia2023foundation} The MACE-MPA-0 (medium) model, trained on an expanded dataset, achieves state-of-the-art accuracy in material property predictions and improved high-pressure stability.\cite{batatia2023foundation} MACE-OMAT-0 (medium) specializes in organic materials tailored to handle the unique bonding and electronic structures of organic compounds.\cite{batatia2023foundation}  Overall, MACE foundation models have demonstrated excellent performance in various domains, from small molecules to large extended materials, outperforming alternative machine learning potentials in both accuracy and data efficiency. \cite{kovacs2023evaluation} In this study, we utilize the MACE foundation models, beginning with rigorous benchmarks to evaluate their accuracy and efficiency for our systems. This is achieved by validating MACE MD simulations across different models against reference AIMD. \\
Among the various MACE models evaluated, MACE-MPA-0\_{medium} was found to yield the most accurate and consistent results (see Figs. \ref{fig:5}).

\clearpage

\clearpage
\begin{figure*}[!htb]
\addcontentsline{toc}{figure}{Figure \ref{fig:6}. Calculated RDFs from MACE-MPA-0\_{medium} model compared to reference AIMD}
\centering
   \includegraphics[width=0.99\textwidth,keepaspectratio]{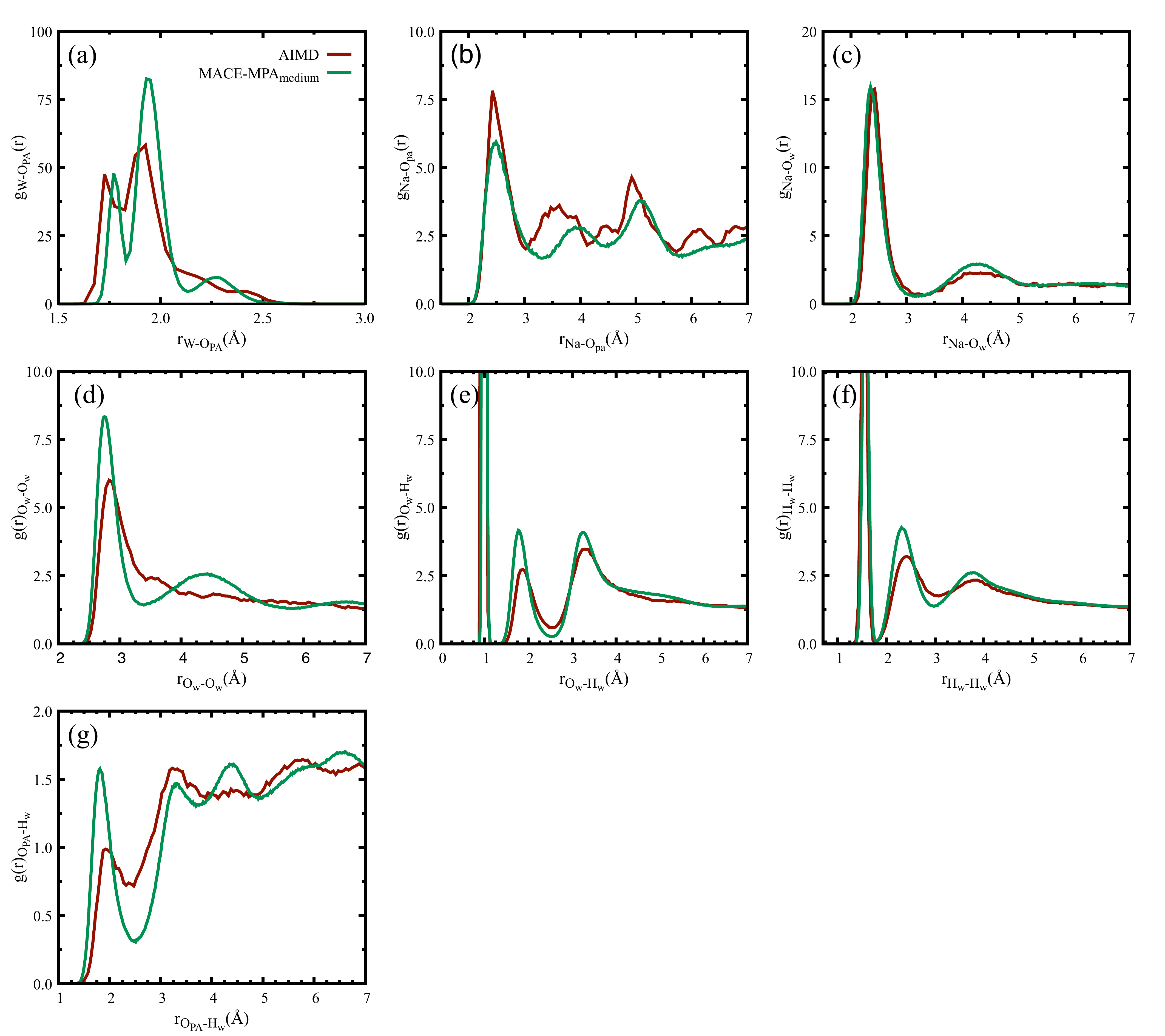}
        \caption{ Calculated RDFs for (a) W--O$_{PA}$ (oxygen of PA), (b) Na--O$_{PA}$, (c) Na--O$_W$ (oxygen of water), (d) O$_W$--O$_W$, (e) O$_W$--H$_W$ (hydrogen of water), (f) H$_W$--H$_W$, and (g) O$_{PA}$--H$_W$ pairs from 1 ns NVT equilibrium simulations for the Na(H$_2$O)@PA system with 110 H$_2$O molecules using MACE-MPA-0\_{medium} model compared to reference AIMD.}    
        \label{fig:6}
\end{figure*}

\clearpage
\begin{figure*}[!htb]
\addcontentsline{toc}{figure}{Figure \ref{fig:7}. MACE MD calculated time evolution of CNs and RDFs in Na@PA }
\centering
   \includegraphics[width=0.99\textwidth,keepaspectratio]{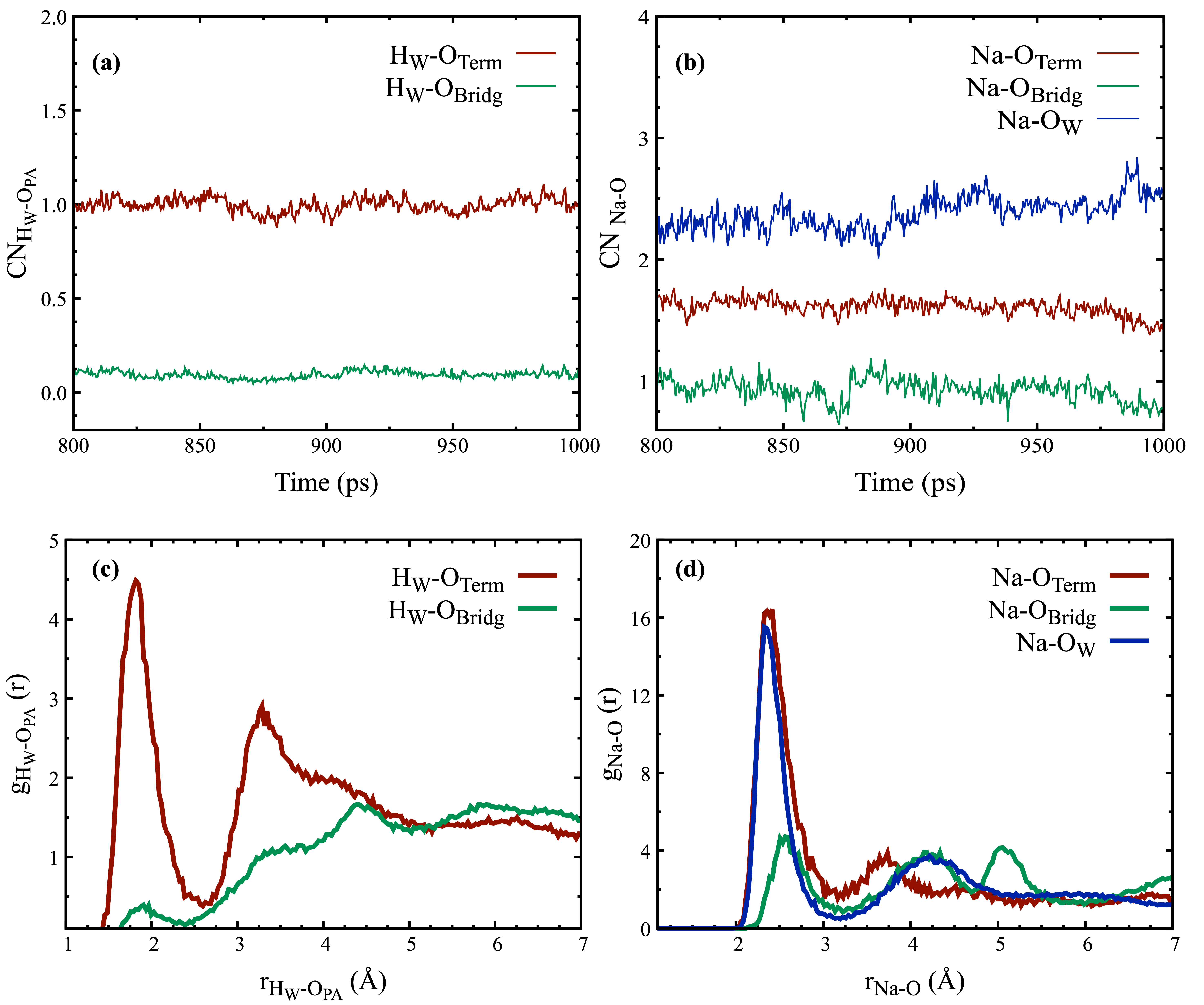}
        \caption{ (top row) Equilibrium MACE MD simulated time evolution of the CNs for (a) hydrogen of the free water molecules to the terminal (brown) and bridging (green) oxygens of the PA in Na@PA and (b) the free Na$^+$ ions to the terminal (brown), and bridging (green) oxygens of the PA in Na@PA as well as water molecules (blue). The bottom panels (c-d) show the corresponding RDFs. See the main text for more details.}    
        \label{fig:7}
\end{figure*}

\clearpage
\begin{figure*}[!htb]
\addcontentsline{toc}{figure}{Figure \ref{fig:8}. MACE MD calculated time evolution of CNs and RDFs in Na(H$_2$O)@PA}
\centering
   \includegraphics[width=0.99\textwidth,keepaspectratio]{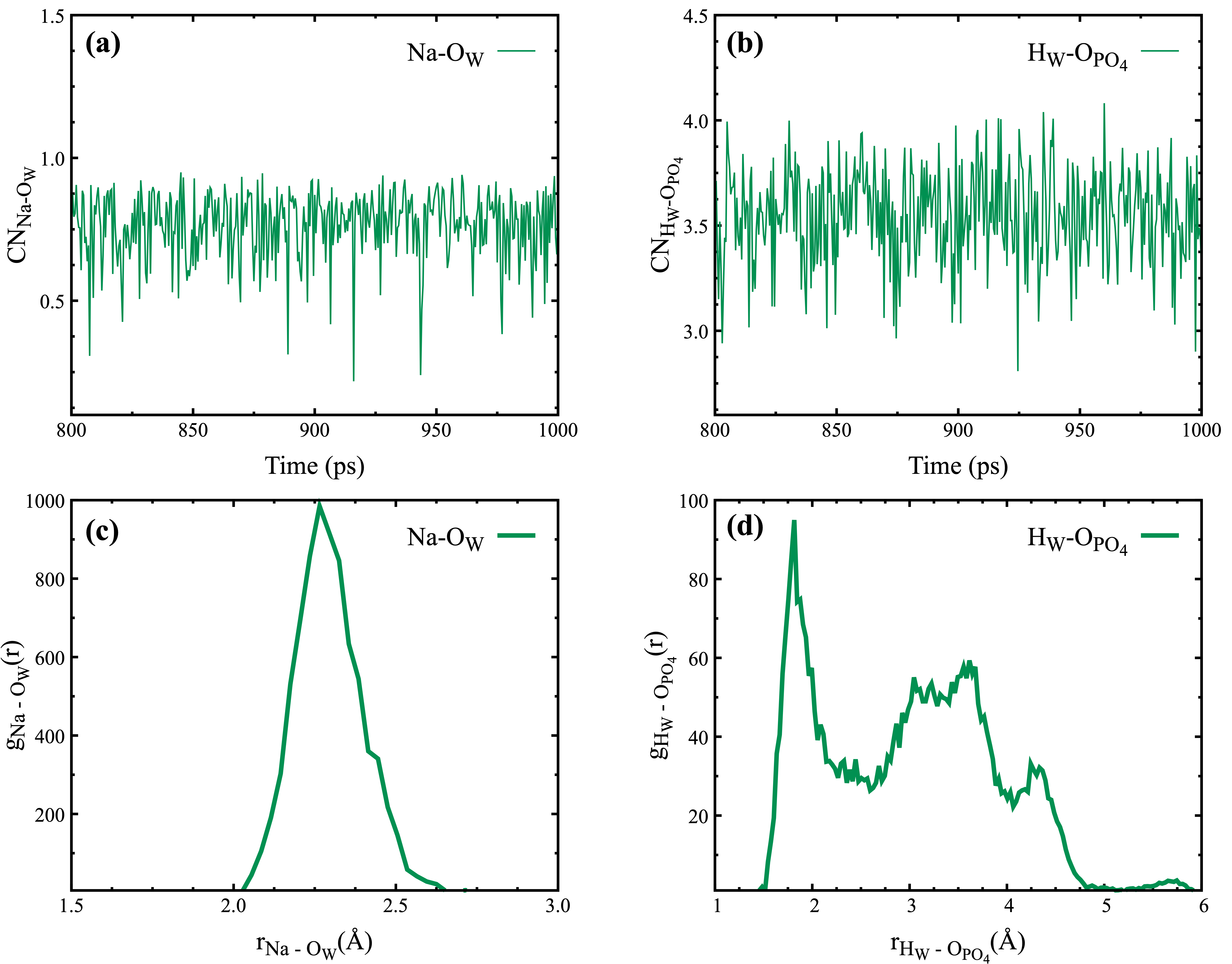}
        \caption{ (top row) Equilibrium MACE MD simulated time evolution of the CNs for the (a) encapsulated Na$^+$ ions to the oxygen of its coordinated water molecule in the PA cavity and (b) hydrogens of the confined water molecule (H$_W$) to the oxygens of the PO$_4$ units. The bottom panels (c-d) show the corresponding RDFs. See the main text for more details.}    
        \label{fig:8}
\end{figure*}

\begin{figure}[!htb]
\addcontentsline{toc}{figure}{Figure \ref{fig:9}. Schematic representation of all CVs.}
\includegraphics [width=0.7\linewidth]{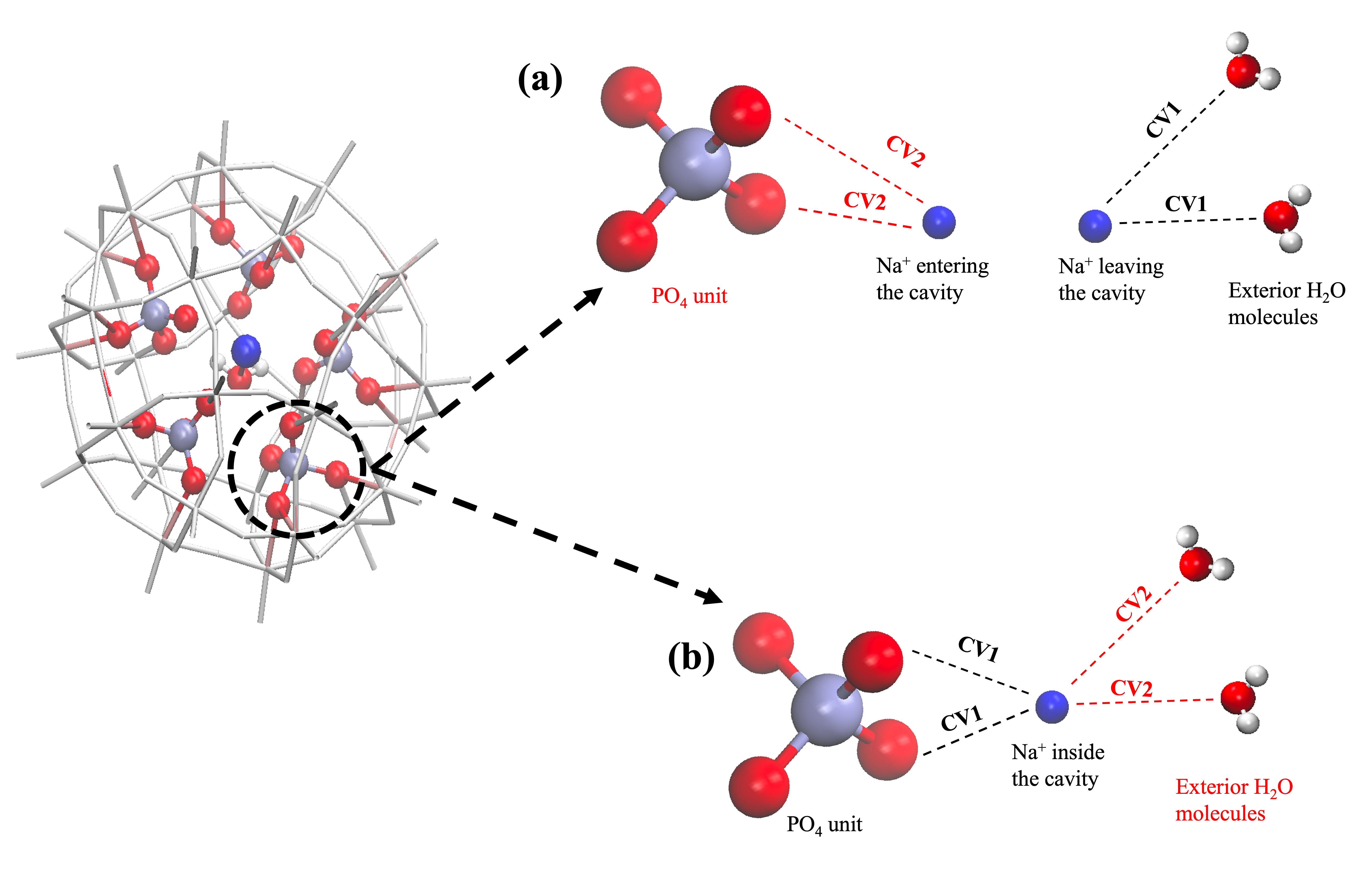}
    \caption{Schematic representation of CVs used in the (a) associative and (b) dissociative mechanisms. Details of the CV definitions and their roles in the respective reaction mechanisms are discussed here and in the main text.}
    \label{fig:9}
\end{figure}

\clearpage
\noindent \textbf{Section S3. Multiple Walker Well-Tempered Metadynamics Simulations}\\
\addcontentsline{toc}{section}{Section S3. Multiple Walker Well-Tempered Metadynamics Simulations}
As mentioned in the main text, metadynamics simulations require a careful selection of appropriate CVs. For our systems, we employed two CVs for each associative and dissociative mechanism (see Fig. \ref{fig:9} for a schematic representation). 
In addition to CV selections, it is equally important to properly fit the switching function for finding an appropriate cutoff distance, r$_0$ , as these parameters critically influence the accuracy and convergence of the free energy surface. To achieve this, we employed RDFs from equilibrium MACE MD simulations (see Fig. \ref{fig:10}). In particular, for the Na@PA and Na(H$_2$O)@PA systems, careful tuning of the switching function and r$_0$ was essential to accurately capture the dynamic coordination changes between the encapsulated Na$^+$ ion, the internal water molecule, and the oxygens of the PO$_4$ unit during the metadynamics simulations. The utilized CVs are governed by the eq. \ref{eq:cv3a} below. 
In the dissociative mechanism for ejecting the confined H$_2$O molecule, first CV is the coordination number between the hydrogen of confined H$_2$O and oxygens of the PO$_4$ unit, with the cutoff distance r$_c$ found to be 2.5~\AA. The second CV, is the coordination between hydrogens of the confined H$_2$O molecule and oxygens of the free H$_2$O molecules at the exterior of the PA where r$_{ij}$ is the distance between the hydrogens of the confined water and oxygens of the free H$_2$O molecules that surrounds the PA, with the cutoff distance r$_c$ is found to be 2.5~\AA. Time evolution of each collective variable is monitored as plotted for each of the walkers in Figs. S11-S14.

\begin{equation}
    S_1 = \sum_{i,j} \left\{\frac{1-(\frac{r_{ij}}{r_c})^{8}}{1-(\frac{r_{ij}}{r_c})^{16}}\right\}
    \label{eq:cv3a}
\end{equation}

\clearpage

\clearpage
\begin{figure*}[!htb]
\addcontentsline{toc}{figure}{Figure \ref{fig:10}. Switching functions and $r_c$ fittings}
\centering
   \includegraphics[width=0.99\textwidth,keepaspectratio]{../figs/Fitt.png}
        \caption{The $r_c$ fittings in switching functions for (a) Na$^+$ to O$_{PA}$  at r$_c$ = 3.2~\AA, (b) Na$^+$ to O$_{W}$ at r$_c$ = 3.0~\AA, (c) H$_W$ to O$_{W}$ at r$_c$ = 2.5~\AA, and (d) H$_W$ to O$_{PA}$ at r$_c$ = 2.5~\AA, see the text for more details.}    
        \label{fig:10}
\end{figure*}

\clearpage
\begin{table}[htbp]
\centering
\addcontentsline{toc}{table}{Table \ref{tab:structural}. Defined CVs and their corresponding cutoff distances}
\small
\renewcommand{\arraystretch}{1.0}
\caption{The defined CVs and their corresponding cutoff distances (r$_c$ in~\AA) used for the various mechanisms. O$_W^f$ and Na$^f$ denote oxygen of the free water molecules and free Na$^+$ ions, respectively.}
\label{tab:structural}
\begin{tabular}{cccccccccccccc}
\toprule
\specialrule{0.1em}{0pt}{0pt}
\multirow{2}{*}{\textbf{System}} & & &\multicolumn{4}{c}{\textbf{Na(H$_2$O)@PA}}&\multicolumn{1}{c}{} \\
%\Xcline{3-9}{1.2pt}
 && CV1 && CV2 &  & r$_c$ (\AA) & &r$_c$ (\AA) & \\

\midrule
\specialrule{0.1em}{0pt}{0pt}
\multirow{1}{*}{\textbf{Ass. Na$^+$}}
   &&Na-O$_W^f$& &Na$^f$-O$_{PO_4}$ &  & 3.2  & & 3.0 \\

   \multirow{1}{*}{\textbf{Dis. Na$^+$}}
   &&Na-O$_{PO_4}$& &Na-O$_W^f$ &  & 3.0  && 3.2 & \\

\multirow{1}{*}{\textbf{Dis. H$_2$O}}
   &&H$_W$-O$_{PO_4}$& &H$_W$-O$_{W}^f$&  & 2.5 & & 2.5  & \\
 \specialrule{0.1em}{0pt}{0pt}
\multirow{2}{*} & & &\multicolumn{4}{c}{\textbf{Na@PA}}&\multicolumn{1}{c}{} &\\
%\Xcline{3-9}{1.2pt}
   \multirow{1}{*}{\textbf{Dis. Na$^+$}}
   & &Na-O$_{PO_4}$& &Na-O$_W^f$ &  & 3.0  && 3.2 & \\
 \specialrule{0.1em}{0pt}{0pt}
\bottomrule
\end{tabular}
\end{table}

\clearpage
\begin{figure*}[!htb]
\addcontentsline{toc}{figure}{Figure \ref{fig:11}. MACE WT-MetaD calculated time evolution of the CVs from all 5 walkers for Na(H$_2$O)@PA in the associative pathway}
\centering
   \includegraphics[width=0.99\textwidth,keepaspectratio]{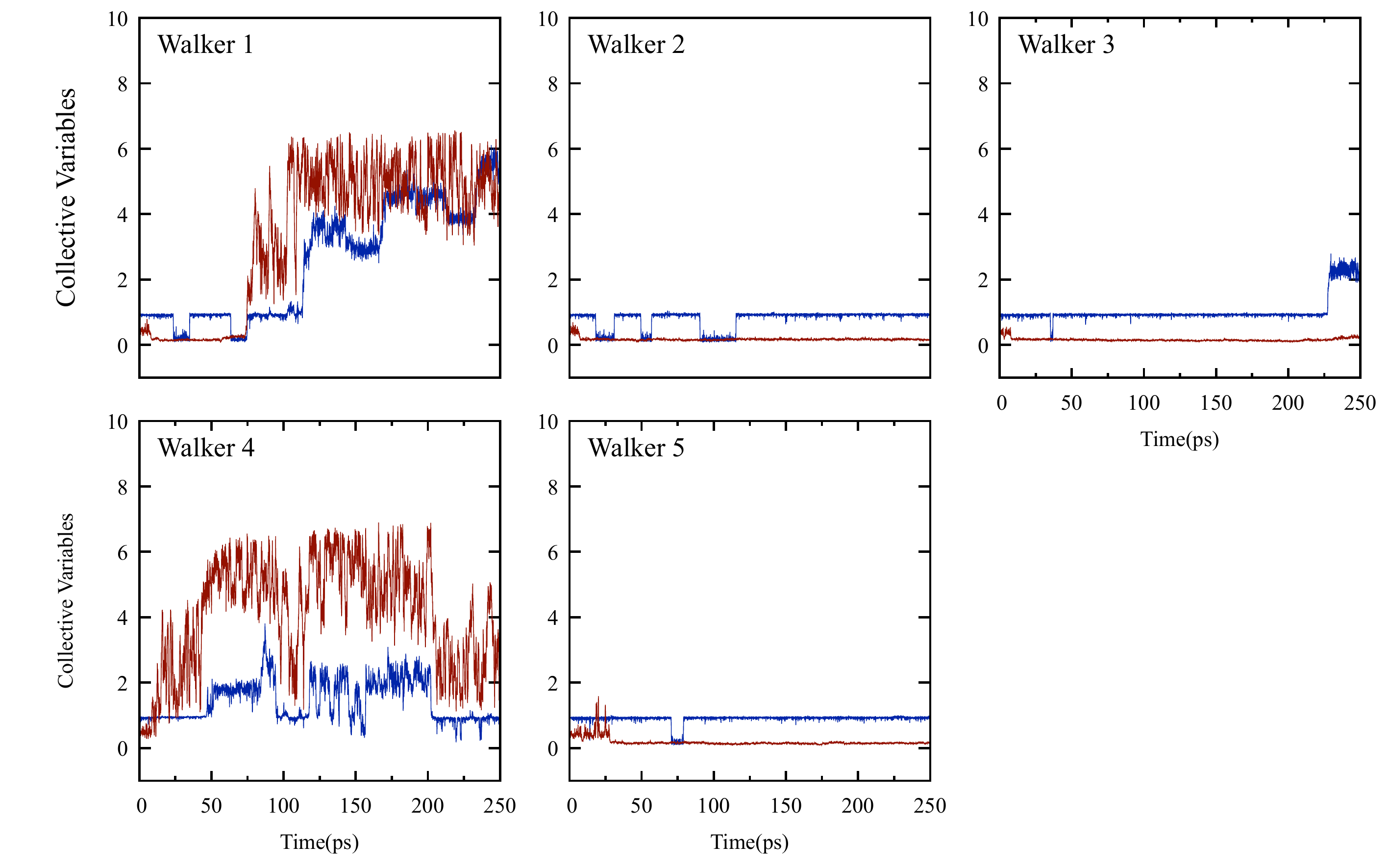}
        \caption{MACE MD calculated time evolution of CVs for 5 walker WT-MetaD simulations performed for the Na(H$_2$O)@PA system in the considered associative pathway.}    
        \label{fig:11}
\end{figure*}

\clearpage
\begin{figure*}[!htb]
\addcontentsline{toc}{figure}{Figure \ref{fig:12}. MACE WT-MetaD calculated time evolution of the CVs from all 5 walkers for Na(H$_2$O)@PA in the dissociative pathway}
\centering
   \includegraphics[width=0.99\textwidth,keepaspectratio]{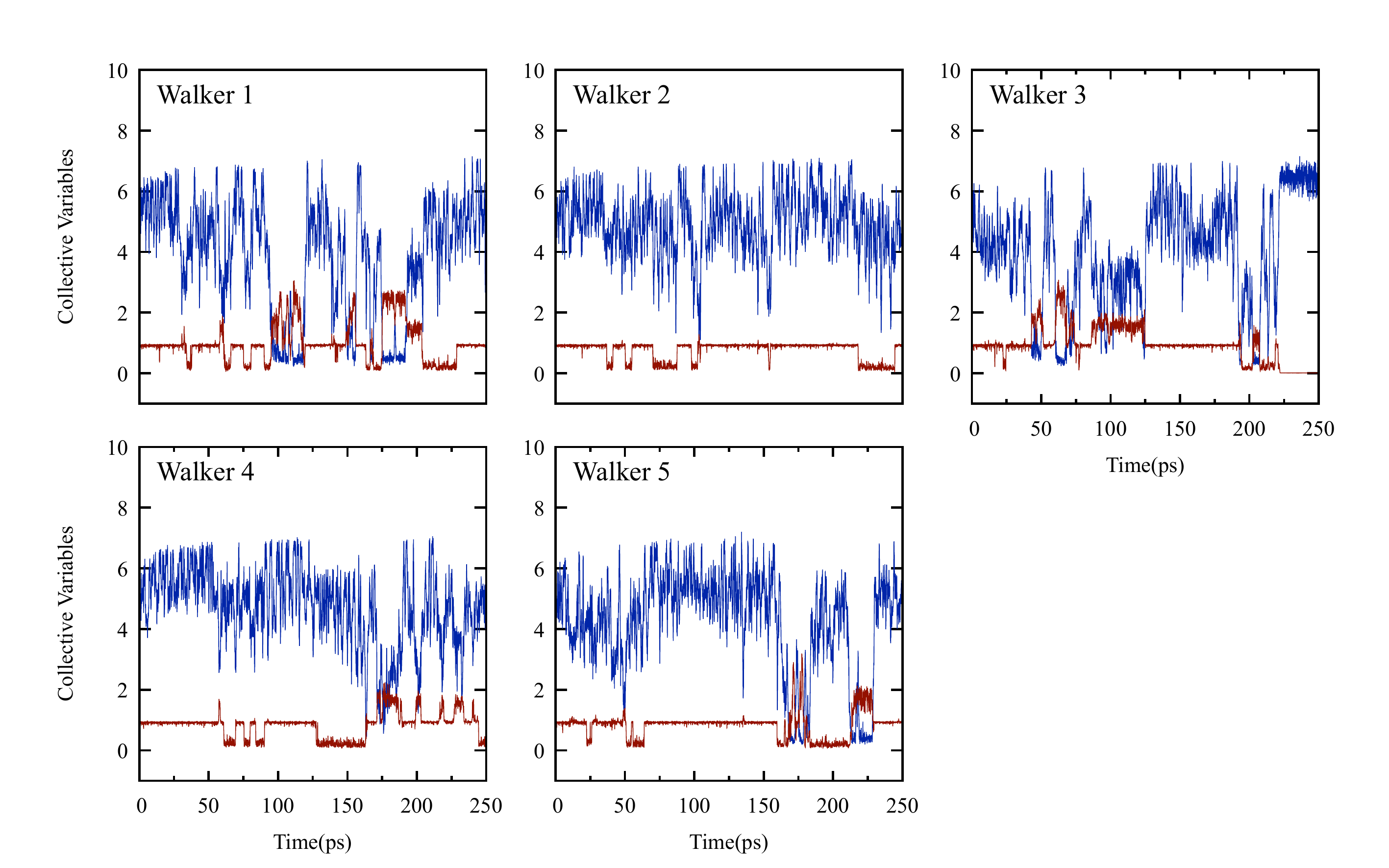}
        \caption{MACE MD calculated time evolution of CVs for 5 walker WT-MetaD simulations performed for the Na(H$_2$O)@PA system in the considered dissociative pathway.}    
        \label{fig:12}
\end{figure*}

\clearpage
\begin{figure*}[!htb]
\addcontentsline{toc}{figure}{Figure \ref{fig:13}. MACE WT-MetaD calculated time evolution of the CVs from all 5 walkers for Na@PA in the dissociative pathway}
\centering
   \includegraphics[width=0.99\textwidth,keepaspectratio]{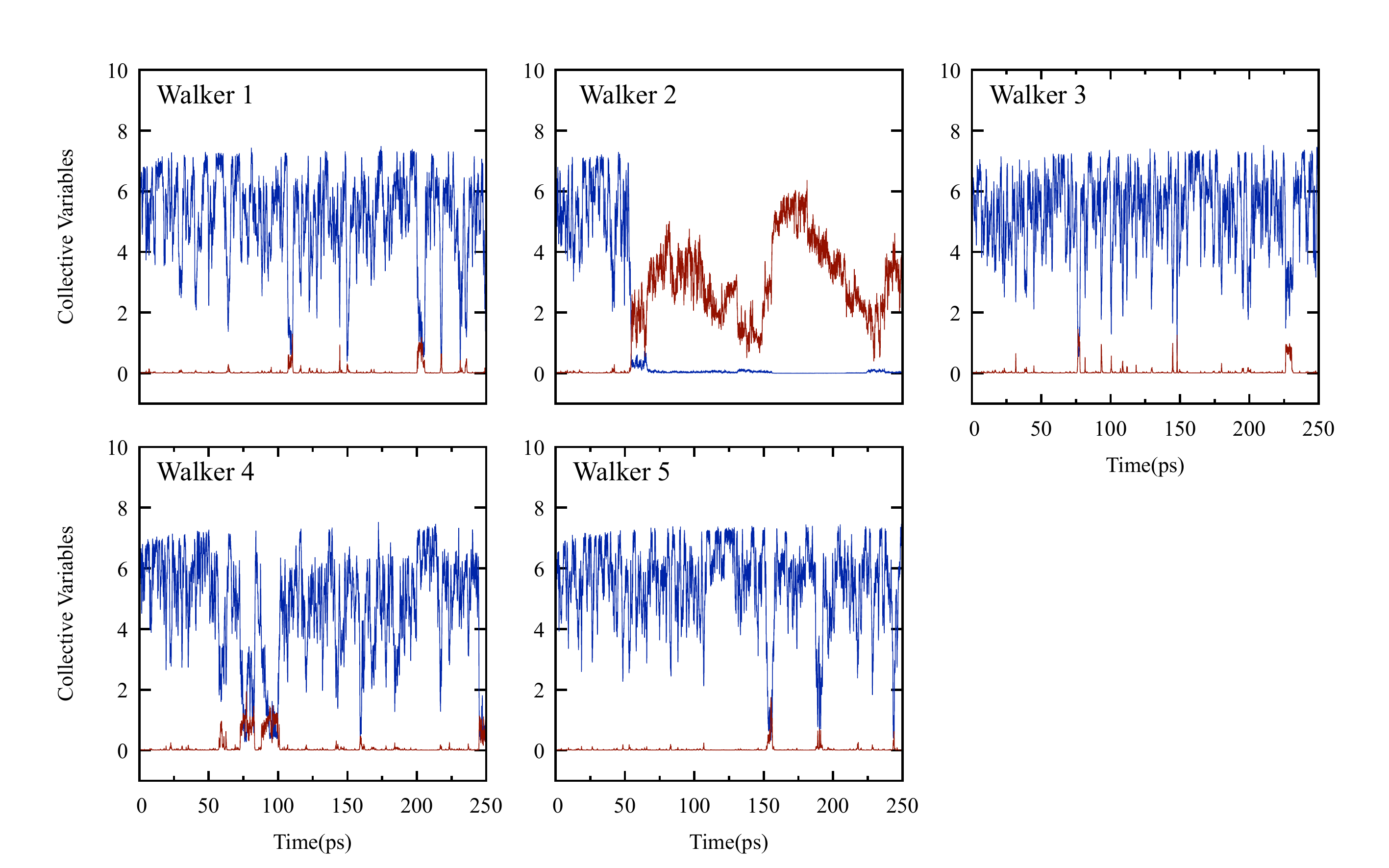}
        \caption{MACE MD calculated time evolution of CVs for 5 walker WT-MetaD simulations performed for the Na@PA system in the considered dissociative pathway.}    
        \label{fig:13}
\end{figure*}

\clearpage
\begin{figure*}[!htb]
\addcontentsline{toc}{figure}{Figure \ref{fig:14}. MACE WT-MetaD calculated time evolution of the CVs from all 5 walkers for Na(H$_2$O)@PA in the dissociative pathway for ejecting H$_2$O}
\centering
   \includegraphics[width=0.99\textwidth,keepaspectratio]{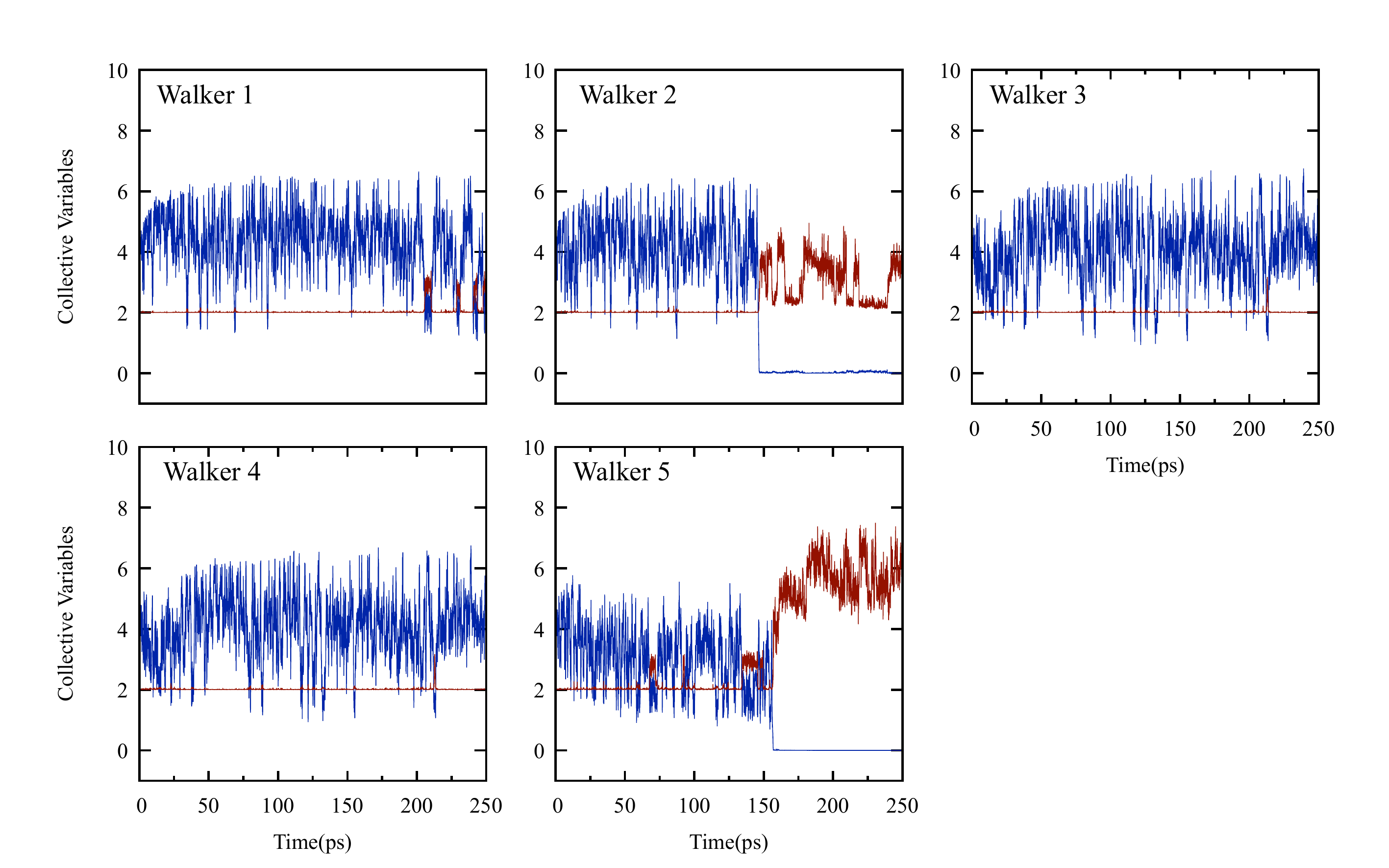}
        \caption{MACE MD calculated time evolution of the CVs for 5 walker WT-MetaD simulations performed for the Na(H$_2$O)@PA system in the considered dissociative pathway for ejecting the confined water molecule.}    
        \label{fig:14}
\end{figure*}

\clearpage
\begin{figure*}[!htb]
\addcontentsline{toc}{figure}{Figure \ref{fig:15}. Fluctuations of window diameters in Na@PA}
\centering
   \includegraphics[width=0.99\textwidth,keepaspectratio]{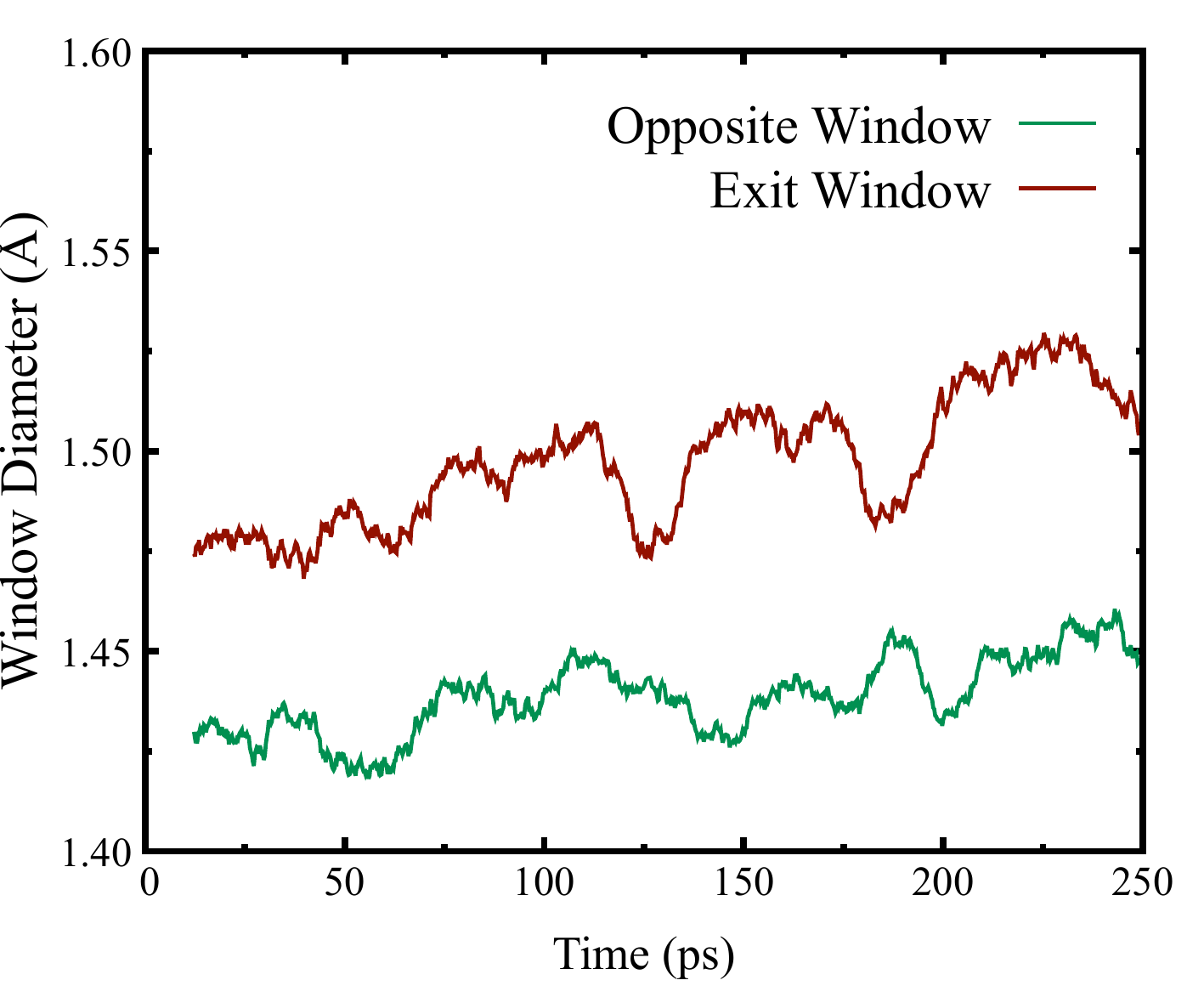}
        \caption{The fluctuation of window diameters in Na@PA, for the considered dissociative pathway for ejecting the encapsulated Na$^+$ ion.}    
        \label{fig:15}
\end{figure*}
\clearpage
\bibliography{../bib}